\documentclass[preprint2]{aastex}

\newcommand{\vsini}{$v$sin$i$}

\newcommand{\teff}{$T_{eff}$}
\newcommand{\logg}{log~$g$}

\newcommand{\ms}{ms$^{-1}$}
\newcommand{\kms}{kms$^{-1}$}
\newcommand{\msol}{M$_{\odot}$}
\newcommand{\me}{M$_{\rm{\oplus}}$}
\newcommand{\rhk}{log$R_{\rm{HK}}$}
\newcommand{\msini}{m~sin~$i$}

\shorttitle{A Uranus-mass planet orbiting HD77338}
\shortauthors{Jenkins et al.}

\begin{document}



\title{A hot Uranus orbiting the super metal-rich star HD77338 and the Metallicity - Mass Connection\thanks{Email: jjenkins@das.uchile.cl. Based on 
observations collected at the La Silla Paranal Observatory, ESO (Chile) with the HARPS spectrograph on the ESO 3.6m telescope, under the program IDs 
079.C-0927, 081.C-0148, 087.C-0368, and 088.C-0662.}}


\author{J.S. Jenkins$^{1,2}$, H.R.A. Jones$^{2}$, M. Tuomi$^{2,3}$, F. Murgas$^{4,5}$, S. Hoyer$^{1}$, M.I. Jones$^{1,6}$, J.R. Barnes$^{2}$, 
Y.V. Pavlenko$^{2,7}$, O. Ivanyuk$^{7}$ , P. Rojo$^{1}$, A. Jord\'an$^{8}$, A.C. Day-Jones$^{1,2}$, M.T. Ruiz$^{1}$ and D.J. Pinfield$^{2}$}
\affil{$^1$Departamento de Astronomia, Universidad de Chile, Camino el Observatorio 1515, Las Condes, Santiago, Chile, Casilla 36-D\\
$^2$Center for Astrophysics, University of Hertfordshire, College Lane Campus, Hatfield, Hertfordshire, UK, AL10 9AB\\
$^3$University of Turku, Tuorla Observatory, Department of Physics and Astronomy, V\"ais\"al\"antie 20, FI-21500, Piikki\"o, Finland\\
$^4$Instituto de Astrof\'isica de Canarias, Via Lactea, E38205, La Laguna, Tenerife, Spain\\
$^5$Departamento de Astrof\'{i}sica, Universidad de La Laguna (ULL),  E-38206 La Laguna, Tenerife, Spain\\
$^6$European Southern Observatory, Casilla 19001, Santiago, Chile\\
$^7$Main Astronomical Observatory of National Academy of Sciences of Ukraine, 27 Zabolotnoho, Kyiv 127, 03680, Ukraine\\
$^8$Departamento de Astronom\'ia y Astrof\'isica, Pontificia Universidad Cat\'olica de Chile, 7820436 Macul, Santiago, Chile}



\begin{abstract}

We announce the discovery of a low-mass planet orbiting the super metal-rich K0V star HD77338 as part of our on-going Calan-Hertfordshire 
Extrasolar Planet Search.  The best fit planet solution has an orbital period of 5.7361$\pm$0.0015~days and with a radial velocity semi-amplitude of only 5.96$\pm$1.74~\ms, 
we find a minimum mass of 15.9$^{+4.7}_{-5.3}$~\me.  The best fit eccentricity from this solution 
is 0.09$^{+0.25}_{-0.09}$, and we find agreement for this data set using a Bayesian analysis and a periodogram analysis.  We measure a metallicity 
for the star of +0.35$\pm$0.06~dex, whereas another recent work (\citealp{trevisan11}) finds +0.47$\pm$0.05~dex.  Thus HD77338$b$ is one 
of the most metal-rich planet host stars known and the most metal-rich star hosting a sub-Neptune mass planet.  We searched for a transit signature of 
HD77338$b$ but none was detected.  We also highlight an emerging 
trend where metallicity and mass seem to correlate at very low masses, a discovery that would be in agreement with the core accretion model of 
planet formation.  The trend appears to show that for Neptune-mass planets and below, higher masses are preferred when the host star is more metal-rich.  
Also a lower boundary is apparent in the super metal-rich regime where there are no very low-mass planets yet discovered in comparison to the sub-solar metallicity regime.  
A Monte Carlo analysis shows that this \emph{low-mass planet desert} is statistically significant with the current sample of 36 planets at $\sim$4.5$\sigma$ level.  
In addition, results from Kepler strengthen the claim for this paucity of the lowest-mass planets in super metal-rich systems.     
Finally, this discovery adds to the growing population of low-mass planets around low-mass and metal-rich stars and shows that very low-mass planets can 
now be discovered with a relatively small number of data points using stable instrumentation.

\end{abstract}


\keywords{stars: fundamental parameters ---  stars: (HD77338) --- stars: rotation --- (stars:) planetary systems}



\section{Introduction}

One of the first correlations announced between two parameters that dealt with exoplanets and their host stars was the 
abundance of heavy elements in the stellar atmospheres.  \citet{gonzalez97} first noted that all three exoplanet host stars known 
at that time were over abundant in iron.  The paper indicated that the metallicity of these stars were all in the super solar regime.  This 
feature has since been studied by a number of authors, most notably by \citet{fischer05} who defined a relationship between 
the host star metallicity ([Fe/H]) and the probability of a star hosting a gas giant planet.

This metallicity bias was one of the major features that helped to confirm that the core accretion scenario of planet formation (e.g. \citealp{ida04}; 
\citealp{mordasini09}), coupled with planetary migration (\citealp{lin86}), appears to be the dominant mechanism to build these planetary systems.  
A further validation of the core accretion mechanism again comes from stellar metallicity since it appears that low-mass Neptunes 
and super-Earths are not predominantly found around metal-rich stars (\citealp{udry06}).

The Calan-Hertfordshire Extrasolar Planet Search (CHEPS) is a program to mainly monitor metal-rich stars to primarily hunt for 
short period planets across a wide range of masses and better constrain the statistics of planets around these stars, whilst also searching 
for planetary transits of bright and nearby stars in the southern hemisphere.  In \citet{jenkins08} we discussed the sample selection 
for the CHEPS, using previous observations made with the ESO-FEROS instrument to measure the chromospheric activity and metallicity 
of a sample of a few hundred stars.  This sample has recently been increased using data discussed in \citet{jenkins11} and current 
metallicity analyses of these stars is still ongoing using our new methods for measuring accurate atomic abundances in stars like 
the Sun (\citealp{pavlenko12}; \citealp{jenkins12}).

In \citet{jenkins09} we published the discovery of an eccentric brown dwarf, or extreme Jovian planet, orbiting the metal-rich star 
HD191760, along with new orbits for three other recently discovered southern hemisphere metal-rich planets.  The CHEPS data has mostly 
been obtained using the HARPS instrument but significant observing time on Coralie has also recently been acquired to pre-select 
interesting new targets for HARPS follow-up (see \citealp{jenkins10b}).

\section{Observations and Reduction}

All the radial velocity data we present in this work were obtained using the HARPS instrument (\citealp{mayor03b}).  From empirical data, HARPS has been shown to be radial velocity stable 
down to the sub-1\ms level (\citealp{pepe11}).  HARPS itself is a fiber-fed 
cross-dispersed echelle spectrograph that employs two 1$''$ fibers, which in simultaneous Thorium-Argon (ThAr) mode places one fibre on the source (HD77338 in this case) 
and one feeds the calibration ThAr lamp to monitor the instrumental drift.  The calibration lamp is useful for sub-1\ms nightly stability as the instrument itself is stable 
at the 1\ms level throughout the course of a single night.

We discuss 32 velocity measurements from HARPS in this current work for the star HD77338 that are shown in Table~\ref{tab:velocities}.  We follow the procedure of observing 
our targets over multiple nights in any single run (see \citealp{jenkins09}), which are usually 3 to 5 days in duration, and with a total integration time of 15~minutes to average over 
the strongest p-mode oscillations in these types of stars (\citealp{otoole09}).  

\begin{table}
\center
\caption{HARPS radial velocities for HD77338.}
\label{tab:velocities}
\begin{tabular}{cccc}
\hline
\multicolumn{1}{c}{BJD} & \multicolumn{1}{c}{RV} & \multicolumn{1}{c}{$\sigma_{int}$}  & \multicolumn{1}{c}{$\sigma_{tot}$}  \\ \hline

2453358.7928045 &  -18.11  &  1.66  &  2.67 \\
2453359.7935070 &    -7.53  & 1.89  & 2.83 \\
2453360.8256229 &  -4.32  & 1.61  & 2.65 \\
2454058.8168860 &  -15.45  &  0.49  &  2.16 \\
2454098.6739885 &  -19.44  &  0.71  &  2.22 \\
2454103.7481653 &  -16.52  &  0.73  &  2.15 \\
2454124.6791307 &   -9.52  & 0.46  & 2.15 \\
2454125.7692682 &  -15.18  &  0.55  &  2.17 \\
2454133.7288993 &  -12.63  &  0.55  &  2.17 \\
2454161.6324334 &  -15.91  &  1.09  &  2.37 \\
2454161.6593658 &  -18.24  &  0.46  &  2.15 \\
2454162.6659105 &  -12.47  &  0.49  &  2.16 \\
2454365.8855969 &  -10.19  &  0.74  &  2.23 \\
2454579.6605011 &  -22.23  &  0.78  &  2.24 \\
2454580.6290112 &  -17.33  &  0.72  &  2.22 \\
2454581.5835519 &   -7.64  & 0.65 &  2.20 \\
2454725.8665876 &  -10.20  &  0.82  &  2.25 \\
2454726.8663334 &  -10.79  &  0.72  &  2.22 \\
2454726.9022721 &  -8.34  &   0.58  &  2.18 \\
2454727.8587883 &  -14.39  &  2.09  &  2.96 \\
2455650.5916958 &  -10.35  &  0.65  &  2.20 \\
2455651.6037866 &  -13.69  &  0.75  &  2.23 \\
2455883.7133405 &  -11.57  &  0.76  &  2.24 \\
2455883.8696951 &  -12.36  &  0.59  &  2.18 \\
2455885.7293282 &  -12.72  &  1.03  &  2.34 \\
2455885.8583195 &  -10.69  &  1.02  &  2.33 \\
2455992.5489188  &   -9.95  &  0.48  &  2.16 \\
2455992.7888091  &   -7.77  &  0.65  &  2.20 \\
2455993.5357474  &   -5.27  &  0.48  &  2.15 \\
2455993.7774660  &   -6.55  &  0.56  &  2.17 \\
2455994.5078761  &   -7.82  &  0.60  &  2.18 \\
2455994.7518050 &  -10.03  &  0.68  &  2.21 \\

\hline
\end{tabular}

\medskip

The $\sigma_{int}$ and $\sigma_{tot}$ are the uncertainties on the points 
before and after adding the estimated activity jitter value shown in 
Table~\ref{tab:values} in quadrature.

\end{table}

Our latest data were reduced and velocities extracted in near real time at the telescope using the most up to date version of the HARPS-DRS (\citealp{pepe04}).  
However, our older data, which were acquired before the latest version of the software was installed on site, were re-reduced offline in the ESO Vitacura offices 
using the latest version of the DRS.  This software performs all functions necessary to fully reduce and analyse a radial velocity timeseries, returning barycentric 
corrected velocities, uncertainties, and a suite of diagnostics to test the line stability of the data set.  All activities we analyse were extracted using a custom 
code developed following our previous experience with this type of chromospheric activity analysis (e.g. \citealp{jenkins06,jenkins08,jenkins11}).

\section{HD77338}

The star HD77338 is classified as K0IV in the Hipparcos main catalogue (\citealp{perryman97}) and has a parallax of 24.54$\pm$1.06~mas (\citealp{vanleeuwen07}), 
giving rise to a distance of 40.75$\pm$1.76~pc.  It has an apparent $V$-band magnitude of 8.63, meaning its distance modulus is 3.05, and the Johnson 
$B-V$ colour index is 0.833.  

The photometric and astrometric analysis means we can place HD77338 on a HR-diagram and compare its position to that of model isochrones and 
isomass tracks.  However, in order to do this properly, we must first have a handle on the metallicity.  Given that the star is fairly nearby and bright, there have been 
a number of attempts to pin down its metallicity.  

Using FEROS spectroscopy we had previously measured the [Fe/H] of HD77338 to be +0.25$\pm$0.07~dex (\citealp{jenkins08}) clearly clarifying that this star is indeed 
super metal-rich.  

\begin{figure}
\vspace{4.0cm}
\hspace{-4.0cm}
\includegraphics{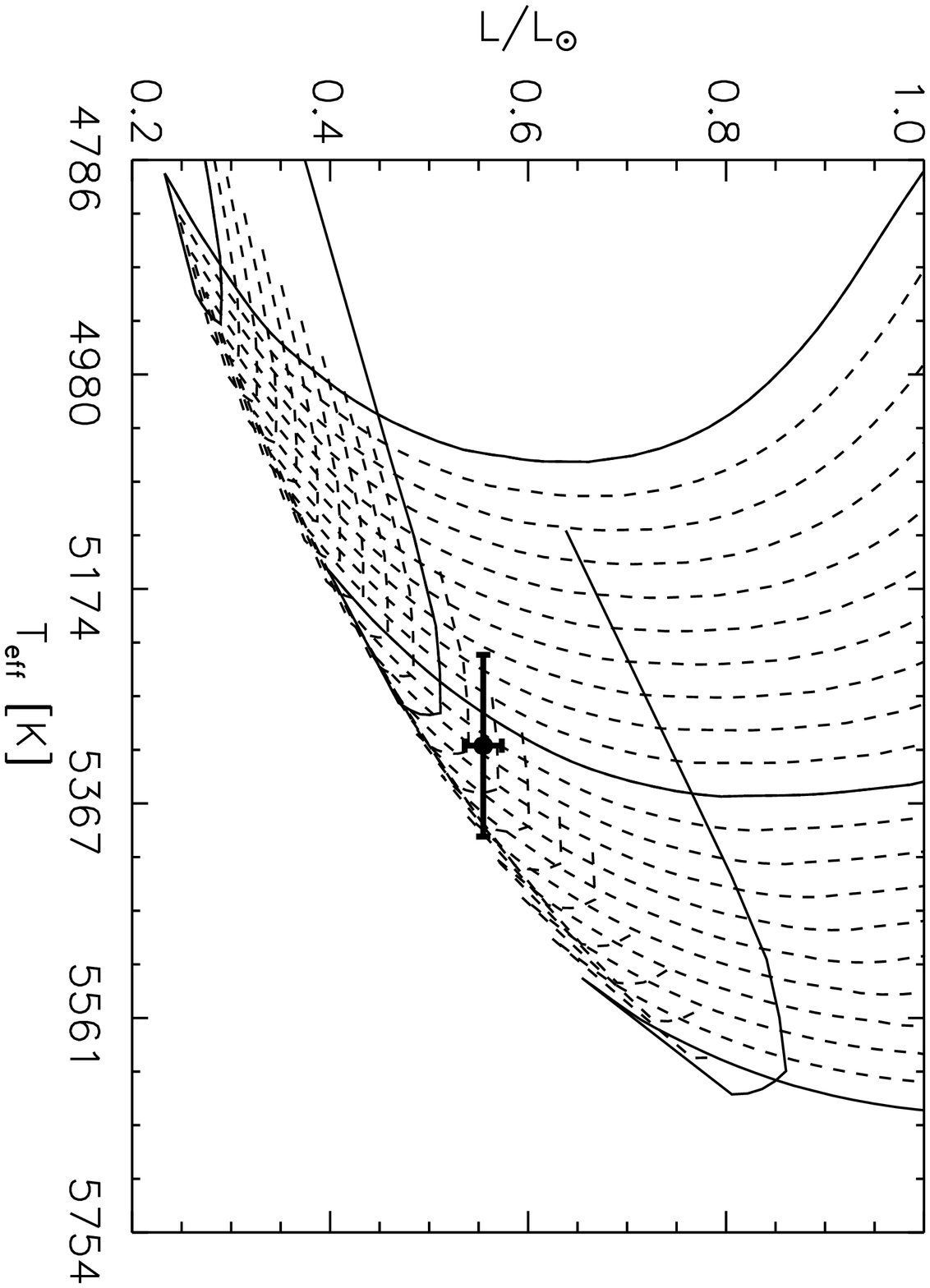}
\vspace{0cm}
\caption{Y2 evolutionary tracks for masses of 1.0, 0.9, and 0.8\msol are shown by the solid curves from top to bottom.  The dashed 
curves are the higher resolution analysis tracks with a step of $\pm$0.01~\msol.  The cross hairs mark the position of HD77338.}
\label{evol}
\end{figure}

Since we now have timeseries data from HARPS we have also measured the metallicity, and other atomic abundances using this spectra.  We normalised each of the 1D spectra 
and combined all the residual fluxes to gain a high S/N and high resolution single HD77338 optical spectrum.  We then applied our new method for measuring self-consistent 
abundances from many optical absorption lines for Fe\sc~i\rm~ and Fe\sc~ii\rm~ independently (\citealp{pavlenko12}) and find a log$N$(Fe\sc~i\rm) of -4.08$\pm$0.02~dex and a 
log$N$(Fe\sc~ii\rm) of -4.07$\pm$0.06~dex.  Although in agreement, the Fe\sc~ii\rm~ abundance has a larger uncertainty due to the lower number of lines used to 
determine this value.  The final abundances for iron and other elements, along with the solar values and the number of lines used in the abundance determination for each are 
shown in Table~\ref{tab:abund}.

\begin{table*}
\center
\caption{Elemental abundances for HD77338.}
\label{tab:abund}
\begin{tabular}{cccccc}
\hline
\multicolumn{1}{c}{Ion}& \multicolumn{1}{c}{log$N$(X)} & \multicolumn{1}{c}{log$N$(X)$_{\odot}$}  & \multicolumn{1}{c}{[X/H]}   & \multicolumn{1}{c}{[X/Fe]}  & \multicolumn{1}{c}{N$_{l}$} \\ \hline

Si\sc~i\rm & -4.00$\pm$0.06  &  -4.45$\pm$0.05$^1$  & 0.45 & 0.10  & 13  \\
Si\sc~ii\rm & -3.78$\pm$0.03  &  -4.45$\pm$0.05$^1$  & 0.67 & 0.32  & 2  \\
Ca\sc~i\rm & -5.25$\pm$0.02  & -5.64$\pm$0.02$^1$  & 0.39  & 0.04  & 12 \\
Ti\sc~i\rm & -6.67$\pm$0.03  &  -6.98$\pm$0.06$^1$  & 0.31 & -0.04  & 29 \\
Ti\sc~ii\rm & -6.76$\pm$0.05 &  -6.98$\pm$0.06$^1$ &  0.22  & -0.13  & 27  \\
Cr\sc~i\rm & -6.04$\pm$0.03 &  -6.33$\pm$0.03$^1$  & 0.29  & -0.06   & 38  \\
Fe\sc~i\rm & -4.08$\pm$0.02 &  -4.42$\pm$0.03$^2$ & 0.34  &  -0.01  &  130 \\
Fe\sc~ii\rm & -4.07$\pm$0.06 &  -4.42$\pm$0.03$^2$  & 0.35  &  0.00  &  18 \\
$\alpha$      & -5.74$\pm$0.09 &  -6.08$\pm$0.11 & 0.34  &  -0.01  & 119 \\
\hline

\end{tabular}

$^{1}$ \citet{grevesse98}, $^{2}$ \citet{pavlenko12}: $\alpha$ = mean of Si\sc~i\rm, Ca\sc~i\rm, Ti\sc~i\rm, 
Ti\sc~ii\rm, and Cr\sc~i\rm.

\end{table*}

Taking these two abundances and subtracting off the solar abundance value of -4.42 (\citealp{pavlenko12}) we find a [Fe/H] of +0.35$\pm$0.06~dex.  This value is in 
agreement with our previous value to within the uncertainties but is even more metal-rich.  It is also in good agreement with other authors, for example \citet{barbuy90} 
find a [Fe/H] of +0.22~dex, \citet{taylor05} find a value of +0.283$\pm$0.051~dex, \citet{cenarro07} quote a value of +0.25$\pm$0.10~dex, and \citet{trevisan11} find 
+0.47$\pm$0.05~dex.  Finally, we can be sure from these analyses that HD77338 is a super metal-rich star.  

In Fig.~\ref{evol} we show the position of HD77338 on a HR-diagram (cross hairs), with \teff on the x-axis and L/L$_{\odot}$ on the y-axis.  Yonsie-Yale (Y2) evolutionary tracks 
are shown (\citealp{demarque04}), and the analysis method we perform is explained in \citet{jenkins09}.  The important point here is that the solid curves represent three 
masses of 0.8, 0.9, and 1.0\msol, from bottom to top, and the dashed curves show the tracks at a higher resolution of 0.01\msol within those three mass regimes.  This 
analysis yields a mass of 0.93$\pm$0.05\msol, a radius of 0.88$\pm$0.04R$_{\odot}$, a \logg~ of 4.52$\pm$0.06~dex, and an age of 3.97$^{+4.23}_{-3.00}$~Gyrs for HD77338.  We note 
that HD77338 is not a subgiant star as labeled in Hipparcos, but is a typical metal-rich K-dwarf star.  All calculated values for HD77338 are shown in Table~\ref{tab:values}.

\begin{table*}
\center
\caption{Stellar parameters for HD77338.}
\label{tab:values}
\begin{tabular}{lll}
\hline
\multicolumn{1}{l}{Parameter}& \multicolumn{1}{l}{HD77338} & \multicolumn{1}{l}{Reference} \\ \hline

RA J2000   (h:m:s)                  & 09$^{\rm{h}}$01$^{\rm{m}}$12$^{\rm{s}}$.494 & \citet{perryman97} \\
Dec J2000 (d:m:s)                  &  -25$^{\rm{o}}$31$'$37.42$''$ & \citet{perryman97} \\
Spectral Type                          & K0IV  &  \citet{perryman97} \\
Spectral Type                          & K0V  &  This Work \\
$B-V$                                     & 0.833 & \citet{perryman97} \\
$V$                                         & 8.63  & \citet{perryman97} \\
distance (pc)                            & 40.75$\pm$1.76 & \citet{vanleeuwen07} \\
log$R'$$_{\rm{HK}}$                   & -5.05 & \citet{jenkins11} \\
Hipparcos $N$$_{\rm{obs}}$             & 172  & \citet{perryman97} \\
Hipparcos $\sigma$                     & 0.013  & \citet{perryman97}  \\
$\Delta$$M_{V}$                        & 0.224  & \citet{jenkins11} \\
$L_{\rm{\star}}$/$L_{\odot}$           & 0.55$\pm$0.02  & This Work \\
$M_{\rm{\star}}$/$M_{\odot}$     & 0.93$\pm$0.05  & This Work \\
$R_{\rm{\star}}$/$R_{\odot}$           & 0.88$\pm$0.04  & This Work \\
$T$$_{\rm{eff}}$ (K)                     & 5370$\pm$82   & This Work \\
$[$Fe/H$]$                              & 0.25$\pm$0.07  & \citet{jenkins08} \\
$[$Fe/H$]$                               & 0.35$\pm$0.06  & This Work \\
\logg                                        & 4.52$\pm$0.06  & This Work \\
U,V,W (km/s)                            & 39.1,-27.7,-24.8 & \citet{jenkins11} \\
$P_{\rm{rot},R'{\rm{HK}}}$ (days)     & 45   & This Work \\
$P_{\rm{rot},v~\rm{sin}~i}$ (days)     & 21.7  & This Work \\
Vrot$_{R'{\rm{HK}}}$ (km/s)      & 1.1  & This Work \\
\vsini (km/s)          & 2.8$\pm$1.5  & \citet{jenkins11}  \\
\vsini (km/s)          & 2.33$\pm$0.05  & This Work \\
Age$_{R'{\rm{HK}}}$ (Gyrs)            & 7.5   & This Work\\
Age (Gyrs)                              & 3.97$^{+4.23}_{-3.00}$   & This Work \\
Jitter - $S_{\rm{MW}}$ (m/s)     & 2.10  & \citet{isaacson10} \\
Jitter - fit (m/s)                     & 1.59   & This Work\\

\hline
\end{tabular}

\end{table*}

\section{Doppler Analysis}

The radial velocity timeseries of data for HD77338 show evidence for a signal with a very short period of only 5.74~days.  The signal is apparent using two separate 
analysis methods that we describe here.

\subsection{Periodogram Analysis}

\begin{figure}
\vspace{5.5cm}
\hspace{-4.0cm}
\includegraphics{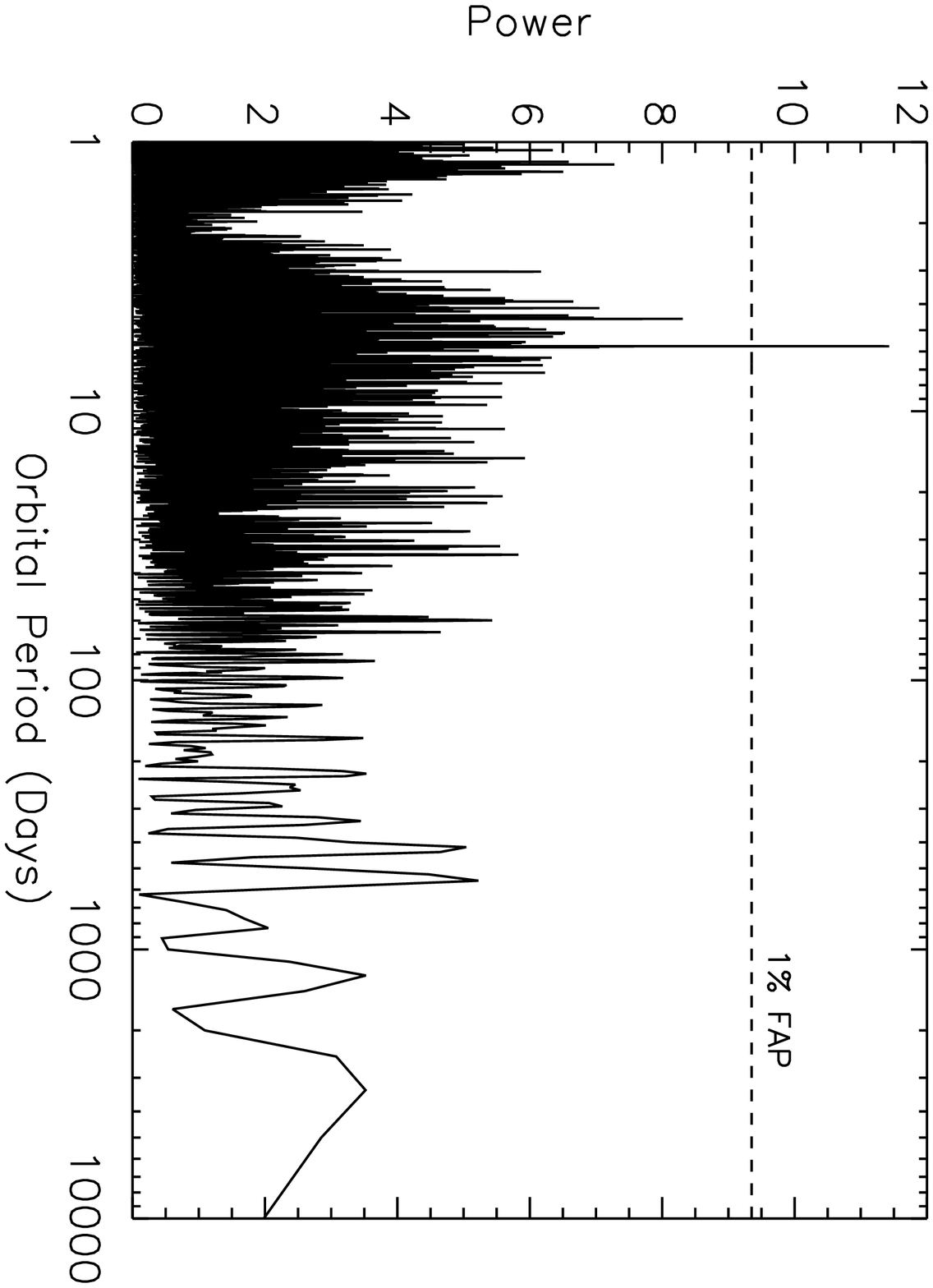}
\includegraphics{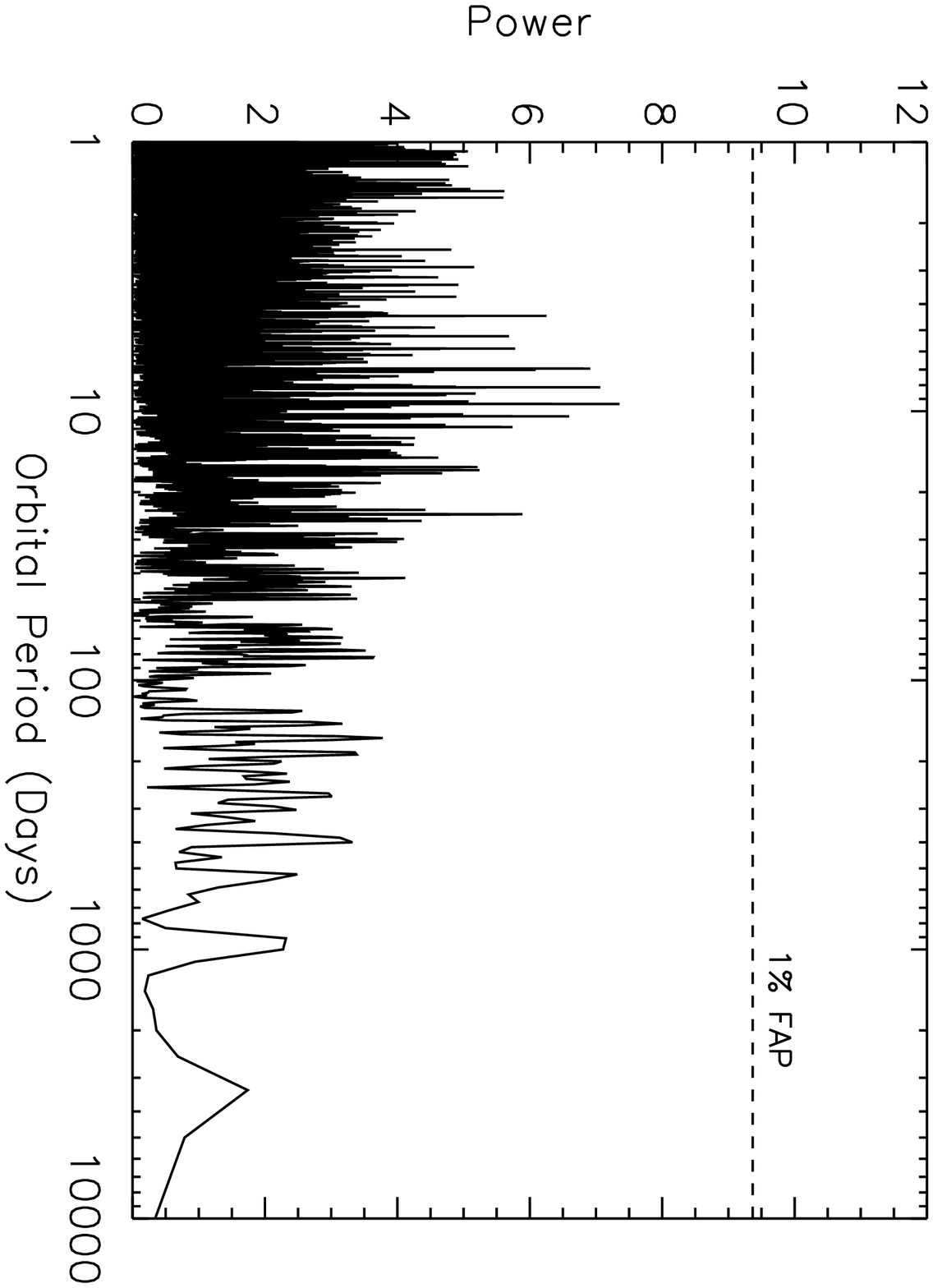}
\vspace{5.5cm}
\caption{Lomb-Scargle periodogram of the HD77338 radial velocity data set is shown in the top plot, with the same analysis performed on the 
residuals to the 1-planet fit shown in the lower plot.  The dashed line boundaries mark the 1\% FAP limit.}
\label{period}
\end{figure}

A periodogram search for strong and stable frequencies in the radial velocity data set for HD77338 reveal a single prominent peak.  
In Fig.~\ref{period} (upper plot) we show the periodogram for the HD77338 data 
and the strongest peak that relates to the 5.74~day Doppler signal is clear.  At present there are still some small peaks around 
this period, but not with a power close to approaching the 5.74~day period.  

Two other peaks are seen in the periodogram, at periods of $\sim$4.5~days and 1.2~days. We have on occasion 
observed this star multiple times per night to help break this aliasing, but it will require further observing runs to suppress the power 
around these frequencies.  In fact in our final three night run, we observed the star twice per night and this significantly reduced the 
power in the 1.2~day signal, indicating this is probably a sampling frequency.  We also note that by looking for power in the sampling, we find some 
power at periods close to one day, further validating the claim that the 1.2~day peak might be a sampling frequency and not a genuine Doppler frequency.  
The strongest peaks in the sampling frequencies are around 35~days, no significant power is found close to the 5.7~day period of our 
planet candidate signature.

We ran a false alarm probability (FAP) test by scrambling the velocities with replacement, but retaining the time-stamps from our observations, to 
test how strong the signal we detect is, in comparison to its emergence being solely through random noise fluctuations.  We ran this bootstrap 
simulation 10000 times, similar to that in \citet{anglada-escude12}, and checked how often the peak power of the scrambled data was higher than or 
equal to the power of the observed data.  This test revealed that 0.01\% of the time the peak power was larger than the observed data set.  This 
gives us confidence that the signal we detect is robust.  The 1\% FAP boundary is shown for reference on the plots.  We also note that the FAP 
registers as 0.02\% if we bin the data points that were observed on the same night.  

We make it clear that the signal peak was already significant with only 27 data points i.e. before our final three night observing run.  At that point the FAP 
registered as 0.02\%, and this led us to obtain two radial velocities per night in our last three night run, since we should sample more than half of 
the planetary candidate orbit in that time.  Since the signal power increased significantly from the addition of these six velocities, we believe it is a 
genuine Doppler reflex motion of the star and not related to a harmonic of our sampling strategy.

The lower plot in Fig.~\ref{period} shows the periodogram of our residuals after subtraction of our best fit solution.  There is no significant power left in the 
data after this single planet fit, an expected result given the relatively small number of data points we have acquired at present.  This also highlights 
how well the Keplerian fit models the current data set as there is no residual power except white noise.  The strongest 
frequency is found around 9.7~days, which could be an emerging planetary signature given the high frequency of multi-planet low-mass 
systems, however as we show below this is likely due to chromospheric activity.

\subsection{Bayesian Search}

We also run a Bayesian search for signals by assuming that 0,1,2,...etc Keplerian signals best 
describe the data we have and find the best fit that conforms to our assigned probability threshold.  We analysed the radial velocities using 
Markov Chain Monte Carlo sampling of the parameter space of each model with a different number of signals.  These samplings were performed using 
the adaptive Metropolis algorithm (\citealp{haario01}; \citealp{tuomi11a,tuomi11b}).  As in Tuomi (2011a, 2011b), we also 
used the sample to assess the relative goodness of the different models by calculating their Bayesian probabilities given the data.  For a 
positive detection of k+1 signals in the data, we required that the posterior probability of a k+1 -signal model was 150 times greater than 
that of a k-signal model (e.g. \citealp{kass95}; \citealp{feroz11}) and that the amplitudes of all the signals were statistically 
distinguishable from zero.

\begin{figure}
\vspace{5.0cm}
\hspace{-4.0cm}
\includegraphics{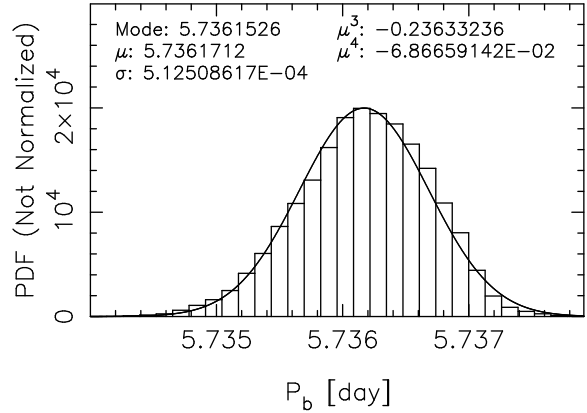}
\vspace{0.5cm}
\caption{Histogram estimating the probability density function of periods from the Bayesian analysis.  Also overplotted by the solid curve is a Gaussian 
curve with the same mean and sigma as the distribution.}
\label{bay_per}
\end{figure}

The Bayesian search we performed using different initial states in the parameter space as starting points of our posterior samplings, 
quickly located a signal with a period of 5.74 days.  This signal satisfied our detection criteria since the probability of no signals is only 6x10$^{-8}$, whereas 
a 1-planet model has a probability of essentially unity.  The samplings did not reveal any other significant signals in the data.  Fig.~\ref{bay_per} shows the probability 
density function (PDF) for the returned periods as a histogram.  The solid curve is a Gaussian profile with the same characteristics as the
histogram, peaking at the measured period, and spread across a confined range of periods. The plot key highlights some of the Gaussian parameters, 
in particular the $\sigma$-width of the distribution shows how well constrained the signal is using this technique, as the range of periods 
here is only at the level of 0.001 days.

\begin{figure}
\vspace{5.0cm}
\hspace{-4.0cm}
\includegraphics{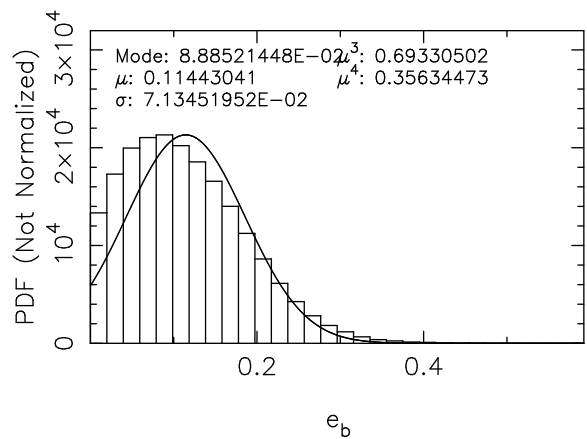}
\vspace{0.5cm}
\caption{Histogram estimating the probability density function of eccentricities from the Bayesian analysis.  Also overplotted by the solid curve is a 
Gaussian distribution with the same mean and sigma as the histogram.}
\label{bay_ecc}
\end{figure}

Fig.~\ref{bay_ecc} shows a similar output PDF but for the possible range of eccentricities.  This histogram does not have a Gaussian shape, more Poissonian, peaking 
around 0.10 or so.  Again a Gaussian profile with parameters from the data is shown for reference.  
The parameters that describe the Gaussian are shown and the $\sigma$ is now at the level of 0.07 in eccentricity.  With a period of less than six days,
putting the planet closer to the star than the tidal dissipation radius, we expect the actual curve is probably more circular, but with the relatively small number of data points, 
we are seeing a more eccentric planetary orbit (\citealp{otoole08}).

To test the plausibility of the non-zero eccentricity model for this system we ran the \citet{lucy71} test on the data.  The test returned a significance, $p$, of 6.7\%, 
showing that the eccentric orbit is not significant at the 5\% threshold level put forward by Lucy \& Sweeney.  Due to this result we also show a model fit that has the 
eccentricity fixed to zero in Table~\ref{tab:orb_params}.

\begin{figure}
\vspace{5.0cm}
\hspace{-4.0cm}
\includegraphics{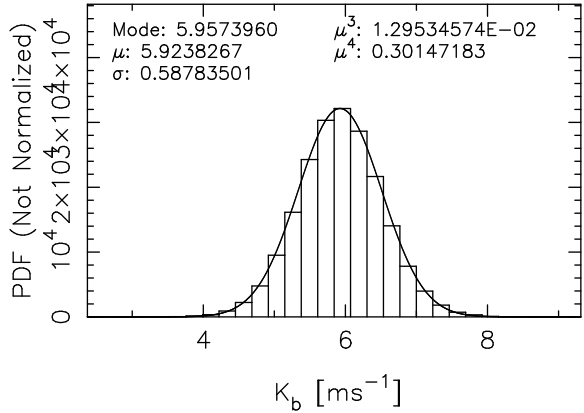}
\vspace{0.5cm}
\caption{Histogram estimating the probability density function of semi-amplitudes for the data from the Bayesian analysis.  Also overplotted by the solid curve is a 
Gaussian distribution with a matching mean and sigma to that of the histogram.}
\label{bay_amp}
\end{figure}

The final PDF we are interested in is shown in Fig.~\ref{bay_amp}, which is for the semi-amplitude of the Doppler signal and shows, in accordance to our 
detection criterion, how it differs significantly from zero.  The peak of the histogram is around 6.0~\ms and the histogram again has a Gaussian 
shape, modeled by the solid curve and from the key we see that the $\sigma$-width is less than 60cms$^{-1}$. From this semi-amplitude and estimated 
stellar mass we directly derive the minimum mass of the orbiting planet, finding a value of 15.9~\me.

\begin{figure}
\vspace{5.0cm}
\hspace{-4.0cm}
\includegraphics{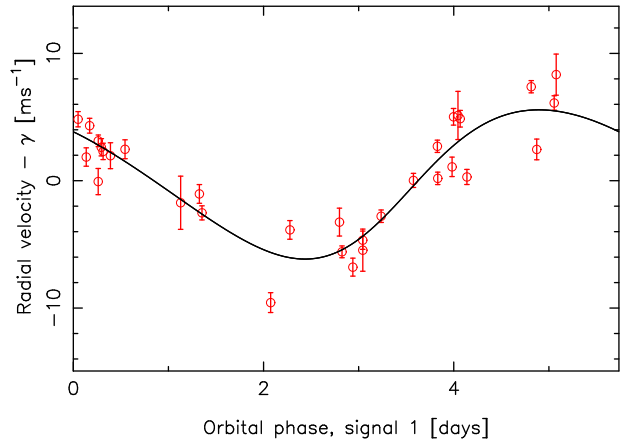}
\vspace{0.5cm}
\caption{Phase folded best fit Keplerian orbital solution found using our Bayesian analysis method is shown by the solid curve.}
\label{bay_phased}
\end{figure}

The final Keplerian curve describing this Doppler signal is shown in Fig.~\ref{bay_phased}, phased to the 5.74~day period of the planet.  It is clear that the data describe the 
periodicity well.  The uncertainties on the data points do not include estimated activity related jitter.  Including the estimated jitter shown in Table~\ref{tab:values} does not 
change any of the final orbital values since we treat the excess noise in the data as another free parameter in our statistical model (see \citealp{tuomi11a}).  Our sampling 
strategy of observing stars over consecutive nights helps to uncover these short periods, giving us efficient access to low-mass planets 
on short period orbits like HD77338$b$.  We also note that the $\gamma$ values we quote are with respect to the local mean of the data in Table.~\ref{tab:velocities}. 

One issue we have to contest with in this data set is the inter-dependency of the velocities we observed in a single night.  On seven of the nights more than one 
radial velocity point was measured, meaning in the strictest sense, the data are not independent.  However, to take this into consideration we have recently developed 
a model that employs an Auto-Regression and Moving Average (ARMA) algorithm to model the noise as an input prior into our full Bayesian model (for details and performance 
of the ARMA model we refer the reader to Tuomi et al. 2012, in preparation).  We ran our ARMA Bayesian model on this data set and find the solution to be statistically similar 
to the ARMA free algorithm.  We find that the moving average components become statistically indistinguishable from zero. Therefore, there are no correlations in the 
noise, at least not large enough for us to detect.

\section{Activity Indicators}

It has been well established that the CCF bisector span (BIS) can correlate with radial velocity timeseries data, indicating that a given frequency detected 
in the velocities is not due to a genuine Doppler motion, but originates from the star itself, from processes that induce line asymmetries in the 
optical spectra of stars (e.g. \citealp{queloz01}; \citealp{santos10}).  Asymmetries can arise from spot patterns rotating across the disk of the star, 
blocking some of the light, and leading to misshapen spectral lines that correlate with the stellar rotation period.  When fixed Gaussian profiles are 
fit to the final CCFs, they do not model the misshapen profile, and so the centroid of the Gaussian ``wobbles'' around the true value, in phase with 
the stars rotation period.

\begin{figure}
\vspace{5.0cm}
\hspace{-4.0cm}
\includegraphics{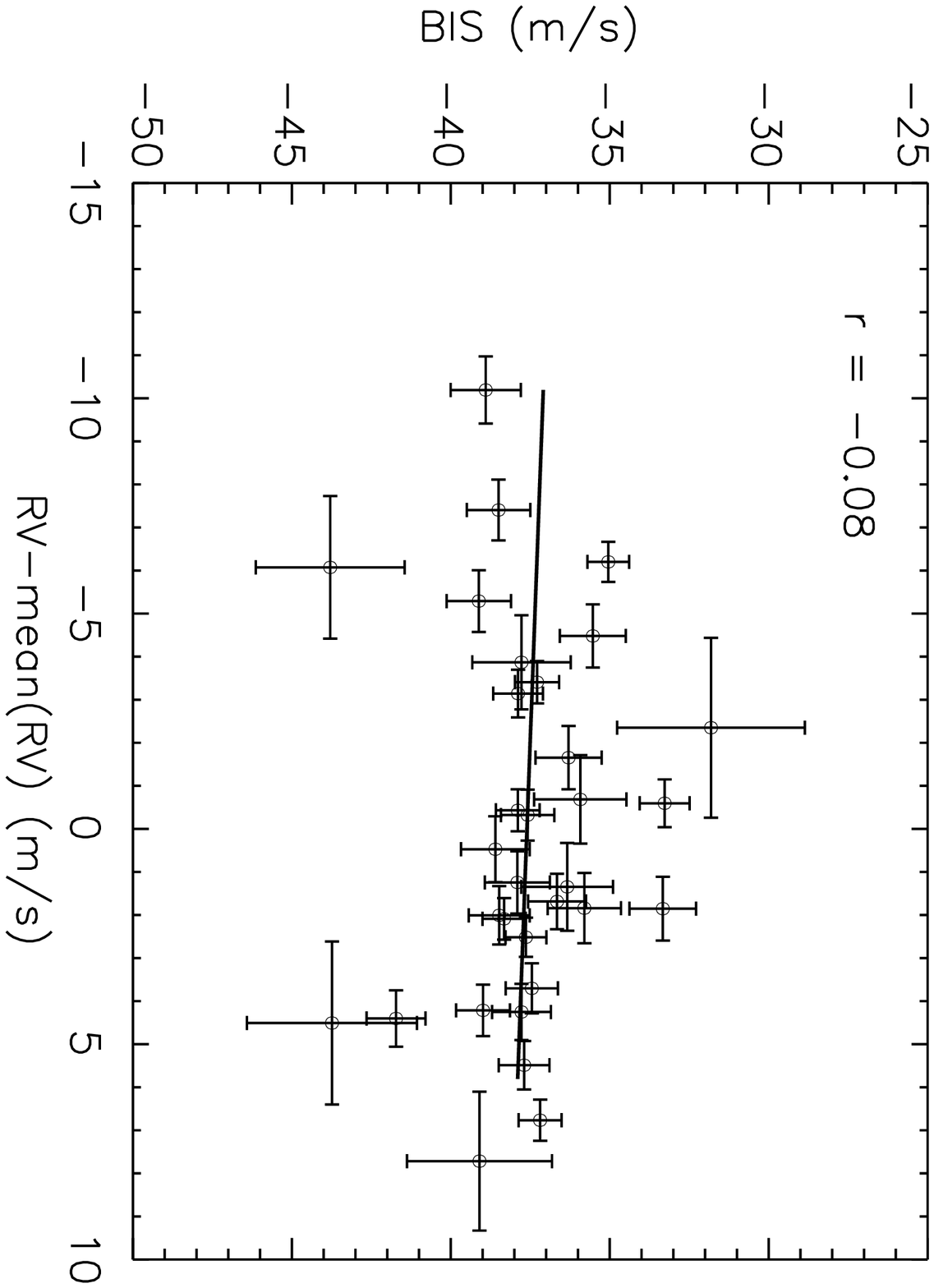}
\includegraphics{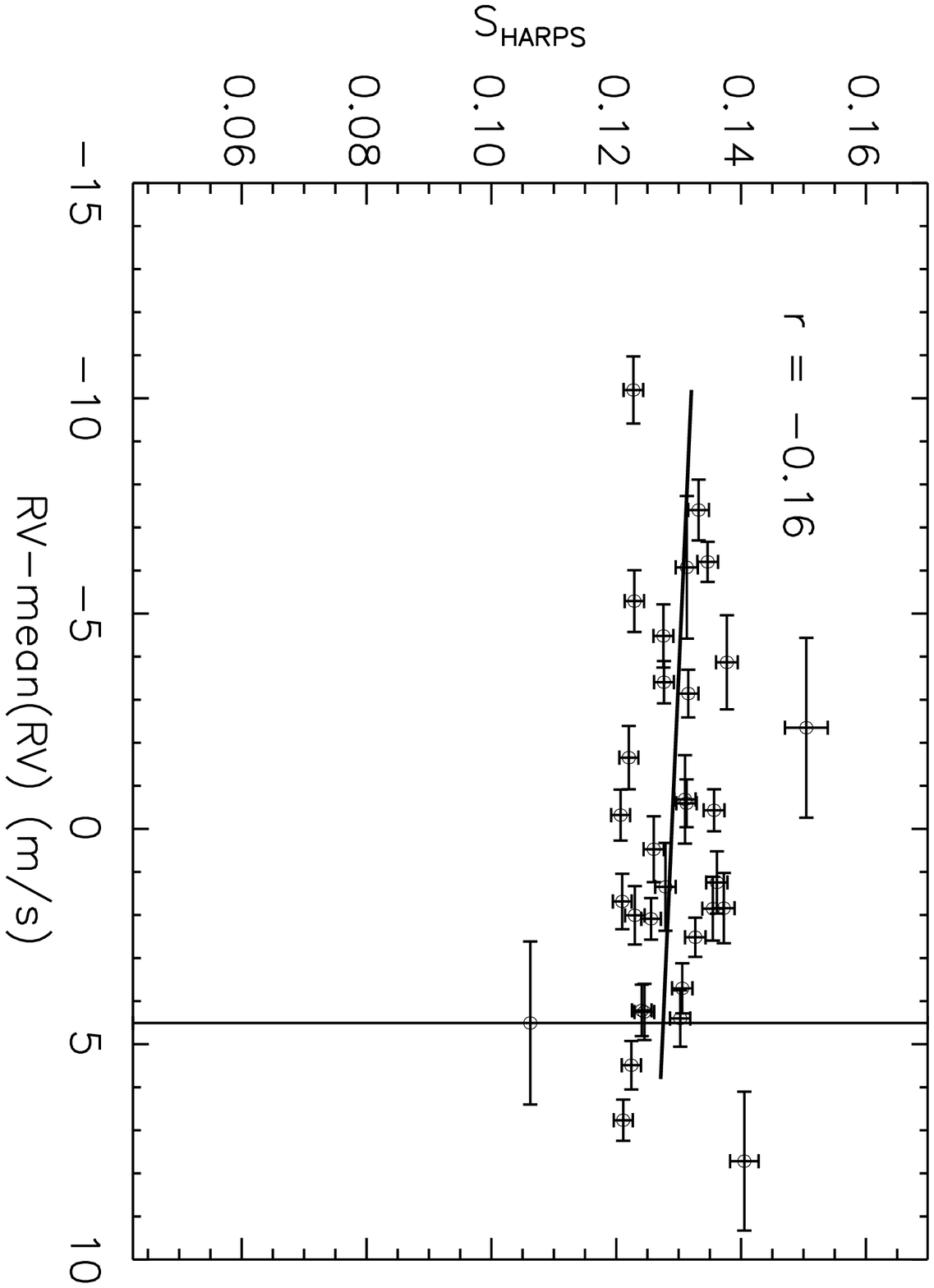}
\vspace{5.2cm}
\caption{Correlations between the radial velocities and the activity indicators that can be drawn from the HARPS reduced spectra.  The top 
panel shows the CCF bisector span against the radial velocities and the lower panel shows the activity $S$ indices for HD77338 as a function 
of the velocities.  The solid lines show the best fit linear correlations between the parameters and the correlation coefficients are highlighted 
in the panels.}
\label{act_ind}
\end{figure}

One test of a true Doppler signal that is not related to stellar activity or velocity fields within the star, is to look for correlations between the BIS and 
the radial velocity timeseries.  In the top panel of Fig.~\ref{act_ind} we show the radial velocities against the BIS for HD77339.  The solid line is the 
best linear fit to the data and reveals that there is no correlation between the two parameters.  To place this on a statistical footing we also show the 
Pearson linear correlation coefficient (r) within the plot, and the value of -0.08 shows that there is no statistical correlation between the radial velocities 
and the CCF BIS values, highlighting that the signal we detect is not due to line asymmetries in the star.  Also a periodogram analysis does not reveal 
any power around a period of 5.7~days.

As mentioned above, magnetically active regions (spots) rotating across the star disk due to the rotation of the star can lead to frequencies appearing 
in the radial velocities.  Given we measured the star to have a \rhk~ of only -5.05, we would not expect the star to be very spotty and also we would not 
expect a rotational period as short as $\sim$6~days.  Also, our \vsini~ measurement is very small (2.33$\pm$0.05~\kms), confirming a longer 
rotational period is preferred for HD77338.  In any case we still test if the chromospheric activity indices correlate with the radial velocities for HD77338.

The lower panel of Fig.~\ref{act_ind} shows the HARPS activity $S$-indices against the radial velocities.  These indices were computed directly from the 
extracted 1D HARPS-DRS spectra, following the methods we have previously laid out in detail in \citet{jenkins06,jenkins08,jenkins11}.  
The plot shows there is no apparent correlation between the two properties.  The solid linear fit has only a small slope and we find the correlation coefficient, r, 
to have a value of -0.16, showing no significant relationship between the activities and the radial velocities.  Again, a periodogram analysis shows no significant 
power anywhere across all periods.  We do note that this fit is unweighted, and the lower activity point with low S/N and large uncertainties is probably pulling 
the slope lower than it should be given the rest of the data set. 

It is also unlikely that rotationally modulated spot patterns could give rise to the detected radial velocity variations since the signal has been stable for at 
least seven years.  Also the measured rotational velocity is very low, meaning a rotation period of less than six days would require an inclination angle of $\sim$18$^{\rm{o}}$, 
assuming a spherical rotating body.  Such a low inclination has probability of only 4.9\%.  The very low chromospheric activity would also argue against a rotational period 
of less than six days, along with the presence of large spot patterns, and given that the star is of K spectral type, we expect a lower contrast between the stellar photosphere 
and spots, in comparison to typical Sun-like G-dwarfs, meaning the radial velocity induced variation from any stable spot patterns would be lower in comparison to 
more Sun-like stars.

\begin{figure}
\vspace{5.0cm}
\hspace{-4.0cm}
\includegraphics{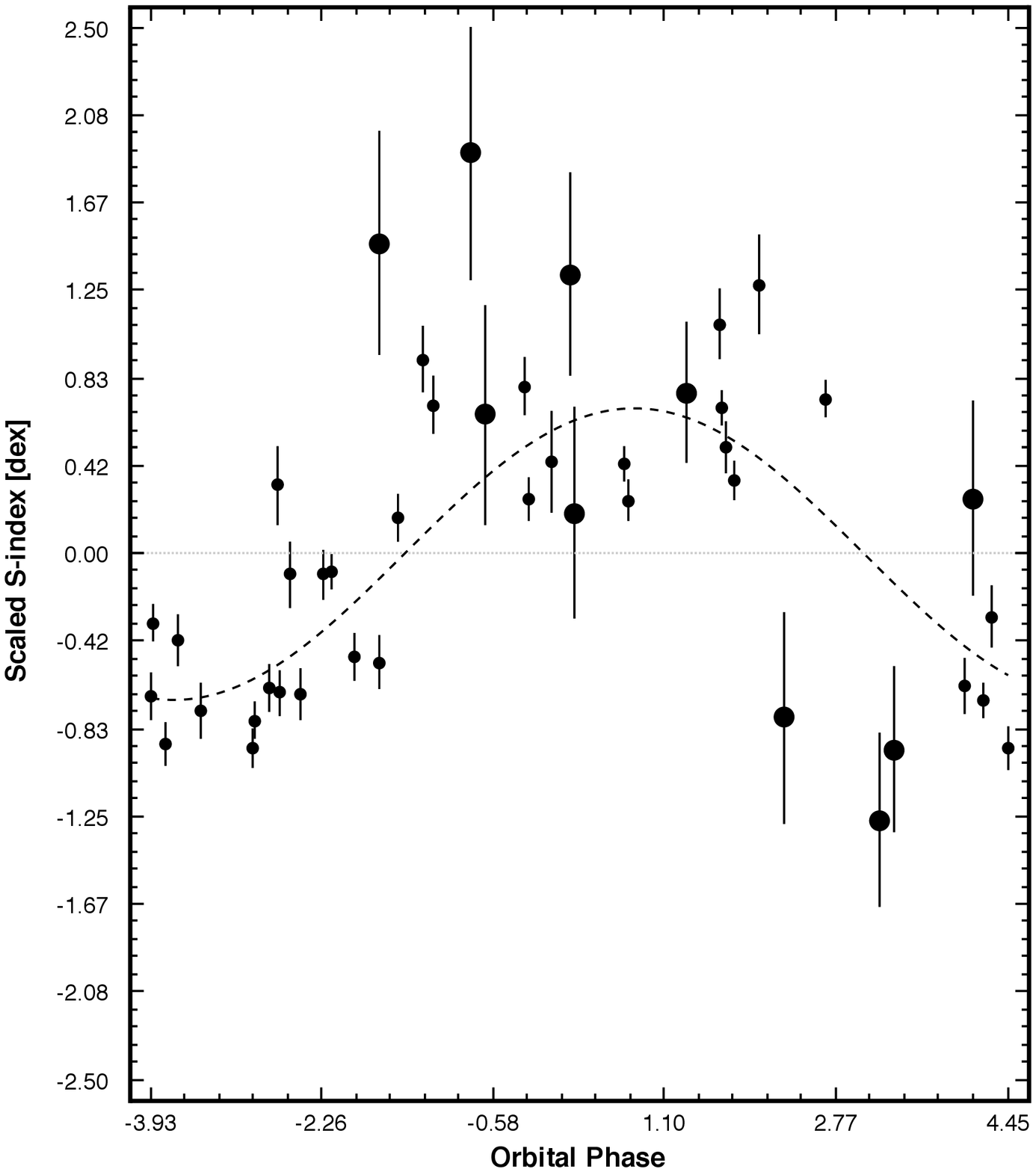}
\vspace{1.3cm}
\caption{The scaled $S$-indices from both HARPS (small uncertainties) and Coralie (large uncertainties) show possible spot modulation 
phased to a period of around 9~days.}
\label{activity_phased}
\end{figure}

A final test we were able to make was done by including the limited number of Coralie activity indices we have for this star.  Inclusion of the Coralie radial velocities does 
not yield any additional confirmation of the reality of the Doppler signal since the uncertainties from our pipeline are larger than the amplitude of the signal, 
however the $S$-indices at this resolution and S/N are accurate enough to help test for spot modulation.  In Fig.~\ref{activity_phased} we show the combined HARPS 
and Coralie $S$-index timeseries phased to the period of the strongest peak in the activity periodogram of 9.001 days (Fig.~\ref{activity_phased}).  HARPS data alone 
favour a 10~day peak signal, with this 9~day signal the second strongest peak, but with the inclusion of Coralie data the 9~day frequency peak is the strongest.  
The period of this possible magnetic modulation is far from the Doppler velocity period yet it is found to be close to where the strongest peak in the residuals of 
the Keplerian fit to the radial velocities is found, as noted above.  This indicates that the peak in the residuals of the best fit to the radial velocities is likely related 
to spot modulation and not a secondary planetary companion.

\begin{table*}
\center
\caption{Maximum a posteriori orbital parameters and their 99\% credibility intervals for HD77338$b$ for unfixed (middle column) and fixed (right column) 
eccentricity.}
\label{tab:orb_params}
\begin{tabular}{ccc}
\hline
\multicolumn{1}{c}{Parameter} & \multicolumn{1}{c}{Model 1} & \multicolumn{1}{c}{Model 2} \\ 
\multicolumn{1}{c}{} & \multicolumn{1}{c}{Unfixed Eccentricity}  & \multicolumn{1}{c}{Fixed Eccentricity} \\\hline

Orbital period $P$ (days)         & 5.7361 [5.7346, 5.7376]     &   5.7361 [5.7345, 5.7374]    \\
Velocity amplitude $K$ (m/s)      & 5.96 [4.22, 7.70]   & 5.96 [4.27, 7.62]  \\
Eccentricity $e$                  & 0.09 [0, 0.34]      &      0.0 [fixed]      \\
$\omega$ (rad)             & 3.9 [0, 2$\pi$]    &   0.0 [fixed]  \\
M$_0$ (rad)                     &  3.5 [0, 2$\pi$]  &   5.8 [0, 2$\pi$]  \\
\msini (M$_{\oplus}$)             & 15.9 [10.6, 20.6]   &  15.5 [11.1, 20.9]    \\
Semimajor axis $a$ (AU)           & 0.0614 [0.0580, 0.0645]     &        0.0611 [0.0581, 0.0642]         \\
$\gamma$ (m/s)                &     -0.64 [-1.81, 0.53]      &     -0.66 [-1.79, 0.36]       \\
rms (m/s)                         & 1.74          &           1.71        \\
N$_{\rm{Obs}}$                     & 32               &   32              \\

\hline
\end{tabular}

The uncertainties on the \msini~ and semimajor axis consider the uncertainties on our 
stellar mass estimate.  The $\gamma$ offset is the value after subtracting off the mean 
of the data set.

\end{table*}

In addition to correlations between the BIS and $S$-indices with the radial velocities, we also test for correlations between the CCF full-width at half maximum 
(FWHM) or the contrast of the CCF with the velocities.  The FWHM and the contrast of the CCF has also been shown to be useful indicators of line variations, 
but at a lower level of significance for these types of stars (\citealp{boisse11}).  Our correlation analyses found no strong correlations between these 
parameters and the velocities, strengthening the case for the reality of a Doppler signal.  The final conclusion that can be drawn from analysis of the properties 
of the CCF and the chromospheric activity values is that the short period signal we have discovered in the radial velocity timeseries of HD77338 is from a genuine 
short period planet orbiting the star, with a low minimum mass that gives rise to the possibility of a planet with a non-negligible rocky mass fraction.  The final 
parameters describing the orbit of HD77338$b$ are shown in Table~\ref{tab:orb_params} for our best fit (top) and with a fixed eccentricity of zero (bottom).

\section{Transit Follow-up}

\begin{figure*}
\vspace{6.0cm}
\hspace{-4.0cm}
\includegraphics{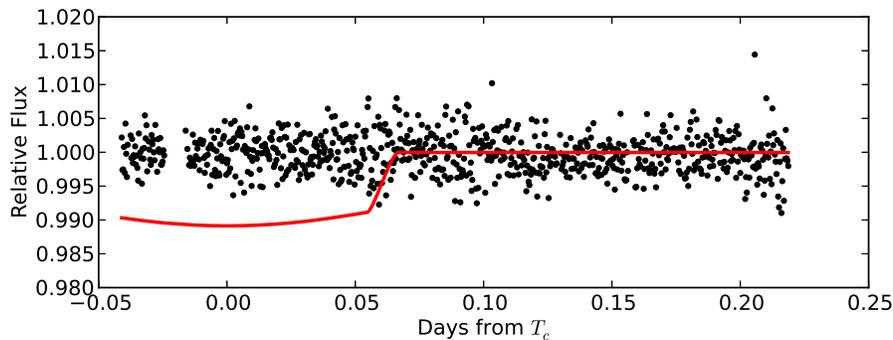}
\vspace{-4.0cm}
\caption{Photometric light curve of HD77338 covering the predicted time the planet would pass in front of the host star.  The 
red curve is the model prediction for how long and how deep the transit should have been if it was observed.}
\label{transit}
\end{figure*}

Short period planets orbiting bright stars can potentially yield transit dips in the light profile of their host stars, and by doing so, can lead to detailed follow-up of atmospheric 
physics, such as those performed on HD209458$b$, HD189733$b$, and GJ1214$b$ (e.g. \citealp{pont08}; \citealp{swain09}; \citealp{beaulieu10}; \citealp{bean11}).  Since 
HD77338 is bright (but not too bright to preclude transit detection using 1-m class telescopes e.g. \citet{jehin10}), having a $V$-band magnitude of 8.63, and the planet has an 
orbital period of less than six days, giving rise to a transit probability of 6.4\%, we performed photometric monitoring around the predicted time of transit from the orbital solution to 
test if HD77338$b$ could be added to this class of bright transiting planets.

We performed the search for the transit of this planet using our current best fit solution for the system parameters, however we suspect a solution with a more circular orbit may be closer to the 
true nature of the HD77338 system. Therefore, assuming a circular orbit and an inclination angle of 90$^{\rm{o}}$, then we can estimate the transit duration of the planet from 

\begin{equation}
\label{transit}
t_{\rm{circ}} = \frac{P}{\pi} \rm arcsin \it \left( \frac{ \sqrt{(R_{\star} + R_{p})^{\rm{2}}} }{a} \right)
\end{equation}

where $P$ is the orbital period, $R_{\star}$ and $R_{p}$ are the stellar and planetary radii respectively, and $a$ again is the semimajor axis.  Assuming a planetary radii commensurate 
with that of Uranus, then the transit duration is predicted to be 2.9~hrs.  Even if the planetary radius is upwards of a Jupiter radius, the transit duration is still estimated to be 2.9~hrs as 
it contributes such a small amount to the net radii in comparison to the star. 

The star was observed on the 25th of March, 2012 and we used the 1m-Cerro-Tololo Inter-American Observatory (CTIO) Telescope, along with the Y4KCam instrument, 
located at the CTIO observatory, in Directors Discretionary Time, to gain the necessary photometric precision to search for a transit of the discovered planet.  
Y4KCam is a 4064x4064 pixel CCD camera with a field-of-view of 20x20 arcmin$^2$ and a pixel scale of 0.289$''$/pixel. The standard readout time of the 
camera is 46s, which we reduce down to 16s by employing a spatial binning of 2x2. 

We employed a strategy of observing the star at integration times of between 7-10 seconds each, with the telescope operating at heavy defocus to get PSFs with a 
typical radius of $\sim$28 pixels.  We obtained 786 data points with typical counts in the range 20'000-25'000, across a time baseline of over 6hrs, crossing the predicted 
center of transit and the predicted egress of planet off the limb of HD77338.  The observed images were then processed using a custom-made pipeline that performs 
the necessary trimming, bias subtraction, and flat-field correction to prepare the images for final analysis.

Fig.~\ref{transit} shows the normalised photometric light curve of HD77338 when the planet was estimated to pass in front of the star.  We find no suggestion of a transit 
around the predicted time of center of transit of JD 2456012.527888.  The rms scatter of our unbinned photometric data is 0.0028 meaning we could significantly detect any 
transit securely with a depth of only 0.008 percent below the continuum level.  

For comparison, the red curve highlights the predicted depth and time of transit for HD77338$b$ given the orbital solution we found from the radial velocity data and a planetary 
radius commensurate with that of Saturn.  Since no 
transit dip was detected we can say that either the inclination of the system is so low that the planet does not pass in front of HD77338 in our line of sight viewing angle, and/or 
the orbital solution we present is not secure enough to pin down a transit detection.  The first of these may be the case and is something we can not do anything about, however 
the second can be remedied with more radial velocity data, which we are still acquiring to search for additional planets in the system.  The Lucy \& Sweeney test we performed also 
indicated that a circular orbit can not be ruled out and therefore we should again search for a transit of HD77338$b$ using the circular orbital solution to test whether or not 
we can detect any occultation of the planet.

\section{Kinematic Motion}

\begin{figure}
\vspace{6.0cm}
\hspace{-4.0cm}
\includegraphics{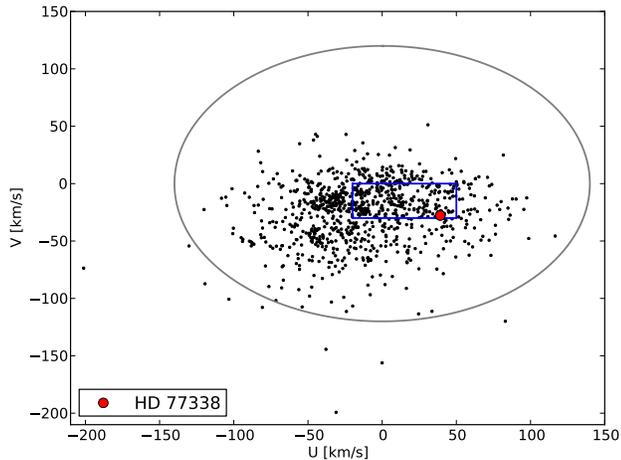}
\vspace{0cm}
\caption{A plot in UV space of the kinematic motion of all stars in Jenkins et al. (2011).  The filled red point marks the position of HD77338.  The solid box and 
solid ellipse mark the regions of the young and old disk, respectively, from \citet{eggen69}.}
\label{kinematics}
\end{figure}

The Hipparcos proper motion of this dwarf star is 41.13$\pm$0.67~mas in $\mu_{\rm{ra}}$ and -270.80$\pm$0.67~mas in $\mu_{\rm{dec}}$, meaning it is classed as a high 
proper motion star.  We used these values to measure the three kinematic velocity components (U,V,W)  (\citealp{jenkins11}), and find velocities of 39.1$\pm$2.0, 
-27.7$\pm$1.8, and -24.8$\pm$1.3~\kms, respectively.  

\begin{figure}
\vspace{5.5cm}
\hspace{-4.0cm}
\includegraphics{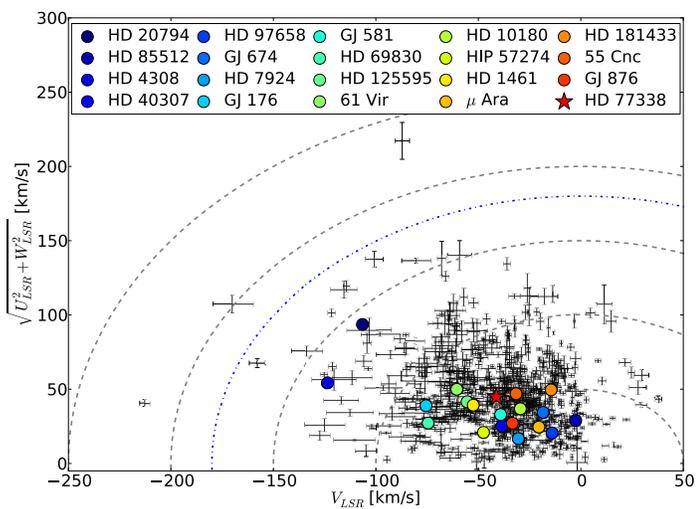}
\vspace{1.0cm}
\caption{A Toomre diagram for the sample of stars in Jenkins et al. (2011) (black points).  The coloured star corresponds to HD77338 and the other coloured circles are all 
radial velocity detected planets with sub-Neptune masses and with colours ordered from the most metal-poor (blue) to the most metal-rich (red).  The data was taken from the Exoplanet 
Database as of February 2012.  The dashed lines mark regions of constant kinetic energy.}
\label{toomre}
\end{figure}

Fig.~\ref{kinematics} shows the distribution of all the stars in Jenkins et al. in the UV-plane, along with the position of HD77338 (filled red circle), the boundaries of 
the young (rectangle), and old (ellipse) disks from \citet{eggen69}.  HD77338 is found close to the boundary between the young and old disk.  Fig.~\ref{toomre} shows 
a Toomre diagram for the same sample and we see here that HD77338 is consistent with either the thin or thick disks, in agreement with the UV-plane kinematics.  
We also show the positions of all other hosts stars that have planets with minimum masses $\le$0.05~M$_{\rm{J}}$ with the colours of the points representing the 
metallicity of the host star, from metal-poor being blue through to red at the metal-rich end. 

In comparison to the other low-mass planet hosts, HD77338 is fairly 
high in the energy plane and very central in the V kinematic velocity plane, along with being the most metal-rich.  As expected, most of the host stars are located around the 
region that contains mostly thin disk stars, with a few probable intermediate and/or thick disk hosts.   However, the two stars HD4308 and HD20794 are found to have kinematics 
that place them well in the thick disk region and encroaching on the halo proxy perimeter.  Also the blue colours of these two show they are 
metal-poor stars which fits in with their kinematics, whereas the hosts with kinematics that place them in the thin disk have a range of metallicities from really metal-poor 
through to super metal-rich.  These two thick disk hosts have very eccentric orbits around the galactic center (Appendix 1; code taken from \citealp{scholz96}) which helps to 
confirm their membership of the 
thick disk since older stellar populations tend to have eccentric orbits and high velocities.  For example, halo stars in general have very eccentric orbits and attain velocities 
above 180~\kms (\citealp{nissen10}) in the Toomre diagram.  If these stars are indeed bonafide members of the thick disk then this helps to show that low-mass planets were 
probably forming early in the formation of the galaxy, if the thick disk was formed through some hierarchal merger or heating process (\citealp{statler88}; \citealp{abadi03}; 
\citealp{villalobos08}; \citealp{brook04}).  However, other recent studies refute the existence of a thick disk being born in this fashion (\citealp{bovy11,bovy12}).

To further test if HD77338 is also a thick disk member we perform an analysis of its orbital motion through the galaxy.  Appendix 2 shows the orbital plots for HD77338 
(top four panels) and we see that it has a fairly eccentric orbit and it travels to $\pm$0.2kpc above and below the plane of the galaxy.  The lower 
panels show the same data for the Sun, a typical thin disk star, and we clearly see the circular orbit around the galaxy.  Also, it only travels half of the distance in Z, above and 
below the galactic plane, in comparison to HD77338.  These points may indicate that HD77338 is actually a metal-rich member of the intermediate/thick disk, rather than the 
thin disk, but thin disk membership can not be ruled out.

The ratio of $\alpha$ elements in comparison to iron can allow one to distinguish between different stellar populations in some cases.  
To parameterize the $\alpha$ abundance of HD77338 we used the $\alpha$ elements shown in Table~\ref{tab:abund}.  In general the measured 
abundances depend on the adopted microturbulent velocity, V$_{t}$ (see \citealp{pavlenko12}).  Using the criteria of the absence of any dependence 
between log N(Fe\sc~i\rm) and log N(Fe\sc~ii\rm) against the excitation potential E$''$ to select the best absorption lines, we found the best fits to the observed profiles 
of the Fe\sc~i\rm~ and Fe\sc~ii\rm~ ions with a V$_{t}$ = 0.5 km/s.  This value was then adopted for the determination of all abundances listed in 
Table~\ref{tab:abund}. 

\begin{equation}
\label{eq:alpha}
[\alpha/Fe] = [\alpha/H] - [Fe/H]
\end{equation}

We used the mean of the elements in Table~\ref{tab:abund} to obtain a 
log$N$($\alpha$) value of -5.74$\pm$0.09~dex for HD77338, and a solar value of -6.08$\pm$0.11~dex, giving rise to a [$\alpha$/H] of +0.34~dex.  By application of 
Eq$^{\rm{n}}$~\ref{eq:alpha} we get the final [$\alpha$/Fe] value of -0.01~dex.

We find no real enhancement of the $\alpha$ abundance for a star of such metallicity.  \citet{bensby05} show abundance trends of such elements and the values of our elements 
agree well with both the thin and thick disk stars for a super metal-rich star.  In fact, Bensby et al. show there there is no significant enhancement of most $\alpha$ elements 
for thick disk metal-rich stars in comparison to metal-rich stars from the thin disk.  The conclusion that can be drawn from the kinematics and abundance analyses is that 
HD77338 is most likely a thin disk star, or in an intermediate region between the thin and thick disks.

\subsection{Elements}

The [X/H] abundances for all the $\alpha$ elements we analysed are much higher than the solar values, in agreement with the super metal-rich nature of HD77338.  
The silicon abundance is significantly enhanced if we consider both the neutral and ionized silicon lines, however as can be seen from Table~\ref{tab:abund} the 
Si\sc~ii\rm~ analysis only employs two useful lines, in comparison to 13 for the Si\sc~i\rm~ abundance.  Recently, \citet{trevisan11} also found a high silicon 
abundance for HD77338 of +0.47$\pm$0.05~dex, in good agreement with the value we find here.

Given that silicon is enhanced in general for thin and thick disk stars in comparison to the amount of iron, it becomes a very important ingredient in planet formation models.  
Large amounts of silicon, and other elements, in the interiors of stars, and by association, the chemical make up of proto-planetary disks, mean a higher disk 
surface density in general.  Changing metallicity, and hence disk surface density, can have a profound impact on planet formation affecting both the masses of planets that 
can form and also by how much they can migrate (see \citealp{mordasini12}).  Metal-rich systems like HD77338 generally give rise to gas giant planets, not lower-mass planets 
like HD77338$b$.  However, population synthesis modeling is a statistical approach and gives general outcomes for the observed population.  It does not preclude the 
formation of very low-mass planets in metal-rich systems, but rather concludes that higher mass planets are more likely around these stars in comparison to stars 
with a lower iron abundance.

The other elements are all found to have values of [X/Fe] commensurate with that of the iron in HD77338.  Indeed, iron has long been used as a strong proxy for the overall metallicity 
([M/H]) of stars like the Sun.  Therefore, for this system we might expect that any core in HD77338$b$ is element heavy and possibly silicate rich.  \citet{sato05} and \citet{fortney06} 
show that the transiting planet HD149026$b$ must be around 2/3 heavy elements by mass, and given that the star is super metal-rich, this points to a connection between the 
abundances of elements in stellar atmospheres and those of their retinue of planets.  A further systematic search for the transit of HD77338$b$ could help to shed light on this issue.

\section{Mass and Metallicity Discussion}

\begin{figure}
\vspace{5.0cm}
\hspace{-4.0cm}
\includegraphics{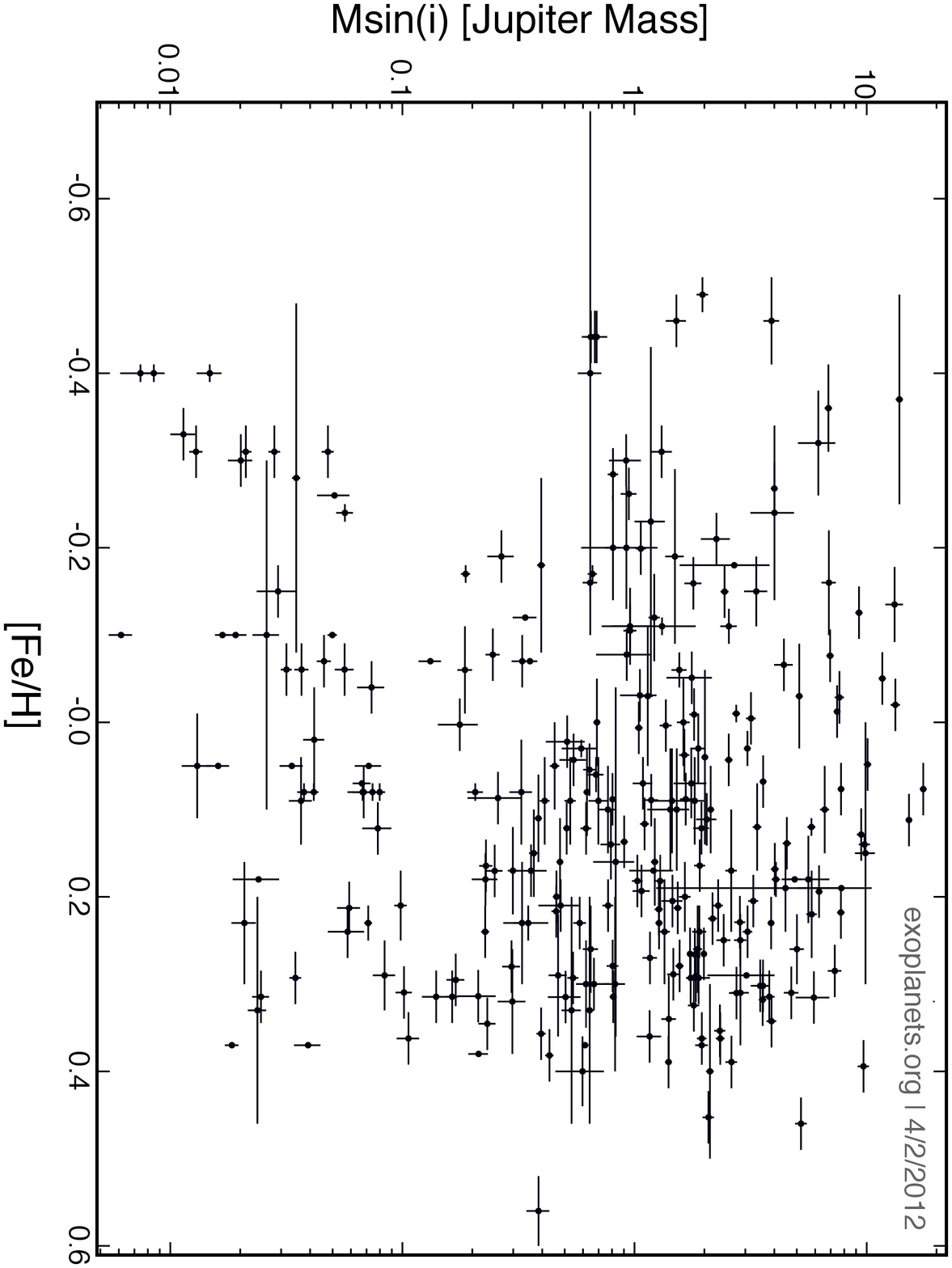}
\vspace{0.5cm}
\caption{Plot of radial velocity detected planets host star metallicity against minimum mass of the planet, taken from the Exoplanet Database (\citealp{wright11}).  
Only planets around dwarf stars were included.}
\label{rv_met_mass}
\end{figure}

HD77338 is one of the most metal-rich stars currently known to host a planet.  If we take the recent \citet{trevisan11} [Fe/H] value of +0.47$\pm$0.05~dex for HD77338 
then only one planet host star has a metallicity higher than this, HD126614~A with a metallicity of +0.56$\pm$0.04~dex (\citealp{howard10}).  However, HD126614~A 
has an M dwarf companion at only 33~AU separation from the host star, meaning HD77338$b$ would be the most metal-rich single star known to host an exoplanet.  
Also, the low-mass nature of HD77338$b$ (\msini~$<$~0.05M$_{\rm{J}}$) helps to populate the low-mass metal-rich bin and shows that metal-rich, very low-mass planets may 
be plentiful.

\begin{table*}
\tiny
\center
\caption{Orbital and stellar parameters for all radial velocity detected host stars with sub-Neptune mass planets.}
\label{tab:planets}
\begin{tabular}{ccccccccccc}
\hline
\multicolumn{1}{c}{Planet} & \multicolumn{1}{c}{\msini} & \multicolumn{1}{c}{Semimajor Axis} & \multicolumn{1}{c}{Period} &
\multicolumn{1}{c}{eccentricity} & \multicolumn{1}{c}{K} &
\multicolumn{1}{c}{$M_{\rm{\star}}$} & \multicolumn{1}{c}{[Fe/H]} & \multicolumn{1}{c}{$V$} & \multicolumn{1}{c}{SpT} & \multicolumn{1}{c}{U,V,W} \\

\multicolumn{1}{c}{} & \multicolumn{1}{c}{M$_{\rm{J}}$} & \multicolumn{1}{c}{AU} & \multicolumn{1}{c}{days} &
\multicolumn{1}{c}{} & \multicolumn{1}{c}{\ms} &
\multicolumn{1}{c}{$M_{\rm{\odot}}$} & \multicolumn{1}{c}{dex} & \multicolumn{1}{c}{mags} & \multicolumn{1}{c}{} & \multicolumn{1}{c}{\kms} \\ \hline

GJ3634$b^1$      &   0.022  &  0.029  &   2.65  &   0.08  &  5.57  &   0.45  &   --  &  11.90   &  M2.5 & -- \\
GJ667C$b^2$      &   0.018  &  0.049  &   7.20  &   0.17  & 3.90   &   0.31  &   -0.59  &  10.22   &  M1.5V & 20,29,-27$^a$ \\
GJ667C$c^2$      &   0.014  &  0.123  &   28.16  &   $<$0.27  &   2.02  &   0.31  &   -0.59  &  10.22   &  M1.5V & 20,29,-27$^a$ \\
HD20794$b^3$   &  0.008   & 0.121   & 18.32   &  0.00   &    0.83   &  0.70   &-0.40   &  4.26    &  G8V & -79,-93,-29$^b$ \\
HD20794$c^3$  &   0.007  &  0.204  &  40.11  &   0.00  &  0.56  &   0.70  & -0.40  &   4.26   &   G8V &-79,-93,-29$^b$  \\
HD20794$d^3$ &    0.015  &  0.350  &  90.31  &  0.00 &   0.85 &    0.70 &  -0.40 &    4.26  &    G8V  & -79,-93,-29$^b$\\
HD85512$b^3$   &  0.011   & 0.260   & 58.43   &  0.11   & 0.77   &  0.69   &-0.33   &  7.67    &  K5V  & 34,11,-5$^b$ \\
HD4308$b^4$    &   0.048  &  0.119  &  15.56  &   0.00  &  4.07  &   0.93  & -0.31  &   6.55   &   G3V & 50,-110,-27$^b$  \\
HD40307$b^5$   &  0.013   & 0.047   &  4.31   &  0.00   & 1.97   &  0.74   &-0.31   &  7.17    &K2.5V & 3,-25,-18$^b$  \\
HD40307$c^5$  &   0.021  &  0.080  &   9.62  &   0.00  &  2.47  &   0.74  & -0.31  &   7.17   & K2.5V & 3,-25,-18$^b$ \\
HD40307$d^5$ &    0.028  &  0.132  &  20.46  &   0.00 &   2.55 &    0.74 &  -0.31 &    7.17  &  K2.5V & 3,-25,-18$^b$ \\
HD97658$b^6$ &    0.020  &  0.080  &   9.50  &   0.13 &   2.36 &    0.75 &  -0.30 &    7.78  &    K1V & -11,-1,-2$^c$ \\
GJ674$b^7$         &  0.035   & 0.039   &  4.69   &  0.20     & 8.70   &  0.35   &-0.28   &  9.36    & M2.5V & -15,-5,-19$^d$ \\
HD7924$b^8$    &   0.029  &  0.057  &   5.40  &   0.17     &  3.87  &   0.83  & -0.15  &   7.18   &   K0V  & 13,-17,-9$^e$ \\
GJ176$b^9$       &    0.026  &  0.066  &   8.78  &   0.00 &   4.12 &    0.49 &  -0.10 &    9.97  &    M2V  & -26,-62,-13$^e$ \\
GJ581$b^{10}$         &  0.050   & 0.041   &  5.37   &  0.03   &12.65   &  0.31   &-0.10   & 10.60    &   M5 & -25,-26,12$^d$ \\
GJ581$c^{10}$          & 0.017    &0.073    &12.92    & 0.07    &3.18    & 0.31  & -0.10  & 10.60     &  M5 &  -25,-26,12$^d$ \\
GJ581$d^{10}$       &    0.019  &  0.218  &  66.64  &   0.25 &   2.16 &    0.31 &  -0.10 &   10.60  &     M5 & -25,-26,12$^d$ \\
GJ581$e^{10}$         &  0.006   & 0.028   &  3.15   &  0.32   & 1.96   &  0.31   &-0.10   & 10.60    &   M5  & -25,-26,12$^d$  \\
BD-082823$b^{11}$ & 0.046    &0.056    & 5.60    & 0.15   &6.50    & 0.74  &  -0.07    & 9.99     & K3V  & --\\
HD69830$b^{4}$    &  0.032   & 0.078   &  8.67   &  0.10   & 3.51   &  0.85   &-0.06   &  5.95    &  K0V  & 29,-61,-10$^b$ \\
HD69830$c^{4}$     & 0.037    &0.185    &31.56    & 0.13    &2.66    & 0.85  & -0.06    & 5.95     & K0V  & 29,-61,-10$^b$ \\
HD125595$b^{12}$   & 0.042    &0.081    & 9.67    & 0.00    &  4.79    & 0.76    &0.02    & 9.03 & K3/K4V  & -34,-42,9$^e$ \\
61Vir$b^{13}$            & 0.016    &0.050    & 4.21    & 0.12    &2.12    & 0.94    &0.05    & 4.87     & G5V  &  -24,-47,-32$^b$ \\
61Vir$c^{13}$          &    0.033  &  0.217  &  38.02  &   0.14 &   2.12 &    0.94 &   0.05 &    4.87  &    G5V  & -24,-47,-32$^b$ \\
HD156668$b^{14}$   &  0.013   & 0.050   &  4.65   &  0.00   & 1.89   &  0.77   & 0.05   &  8.42    &  K3V  & -- \\
HD10180$c^{15}$    &   0.042  &  0.064  &   5.76  &   0.08  &  4.54  &   1.06  &  0.08  &   7.33   &   G1V  &  9,-16,-30$^b$ \\
HD10180$d^{15}$  &     0.038  &  0.129  &  16.36  &   0.14 &   2.93 &    1.06 &   0.08 &    7.33  &    G1V  & 9,-16,-30$^b$ \\
HIP57274$b^{16}$     &  0.037   & 0.071   &  8.14   &  0.19   & 4.64   &  0.73   & 0.09   &  8.96    &   K8  & 2,-34,27$^e$ \\
HD1461$b^{17}$       &  0.024   & 0.064   &  5.77   &  0.14  & 2.70   &  1.03   & 0.18   &  6.60    &  G0V  & -31,-39,-1$^b$ \\
HD215497$b^{18}$   &  0.021   & 0.047   &  3.93   &  0.16   & 2.98   &  0.87   & 0.23   &  9.10    &  K3V  &  -- \\
$\mu$Ara$d^{4}$          &  0.035   & 0.093   &  9.64   &  0.17  & 3.06   &  1.15   & 0.29   &  5.12    &  G3V  & -15,-7,-3$^e$  \\
55Cnc$e^{10}$           & 0.025    &0.015    & 0.74    & 0.06    &5.92    & 0.90    &0.31    & 5.96     & G8V  &  -37,-18,-8$^d$\\
GJ876$d^{19}$            & 0.018    &0.021    & 1.94    & 0.21    &6.56    & 0.32    &0.37  & 10.20     &  M5   &  -13,-20,-12$^d$ \\
GJ876$e^{19}$            & 0.039    &0.333  &124.26    & 0.05   &3.42    & 0.32    &0.37  & 10.20     &  M5  & -13,-20,-12$^d$ \\

\hline
\end{tabular}

\medskip

$^1$\citet{bonfils11}; $^2$\citet{anglada-escude12}; $^3$\citet{pepe11}; $^4$\citet{valenti05}; $^5$\citet{mayor09}; $^6$\citet{henry11}; $^7$\citet{bonfils07}
$^8$\citet{howard09}; $^9$\citet{forveille09}; $^{10}$\citet{vonbraun11}; $^{11}$\citet{hebrard10}; $^{12}$\citet{segransan11}; $^{13}$\citet{vogt10}; $^{14}$\citet{howard11}
$^{15}$\citet{lovis11}; $^{16}$\citet{fischer12}; $^{17}$\citet{rivera10}; $^{18}$\citet{locurto10}; $^{19}$\citet{johnson09} \newline
$^a$\citet{anglada-escude12}; $^b$\citet{holmberg09}; $^c$\citet{marsakov88}; $^d$\citet{montes01}; $^e$\citet{woolley70}

\end{table*}

The high abundance of all the elements we have analysed also indicates that the proto-planetary 
disk left over from the formation of HD77338 was rich in planet building material.  The established preponderance for gas giant planets to favour metal-rich stars 
(\citealp{gonzalez97}; \citealp{fischer05}; \citealp{sousa11}) can be seen in Fig.~\ref{rv_met_mass}.  We also see 
that metal-rich stars tend to cover the entire phase space of planetary masses.  Above a host star metallicity of $\sim$0.2~dex there are 
planets covering the whole regime from low-masses to high-masses, in dense clusters.  However, for lower metallicity systems, particularly at sub-solar metallicities, there 
are regions free from any planet detections, or at least less densely packed.  This shows why metal-rich stars are so highly prized in the hunt for exoplanets and for better 
understanding the nature of planet formation and migration.

\begin{figure}
\vspace{7.0cm}
\hspace{-4.0cm}
\includegraphics{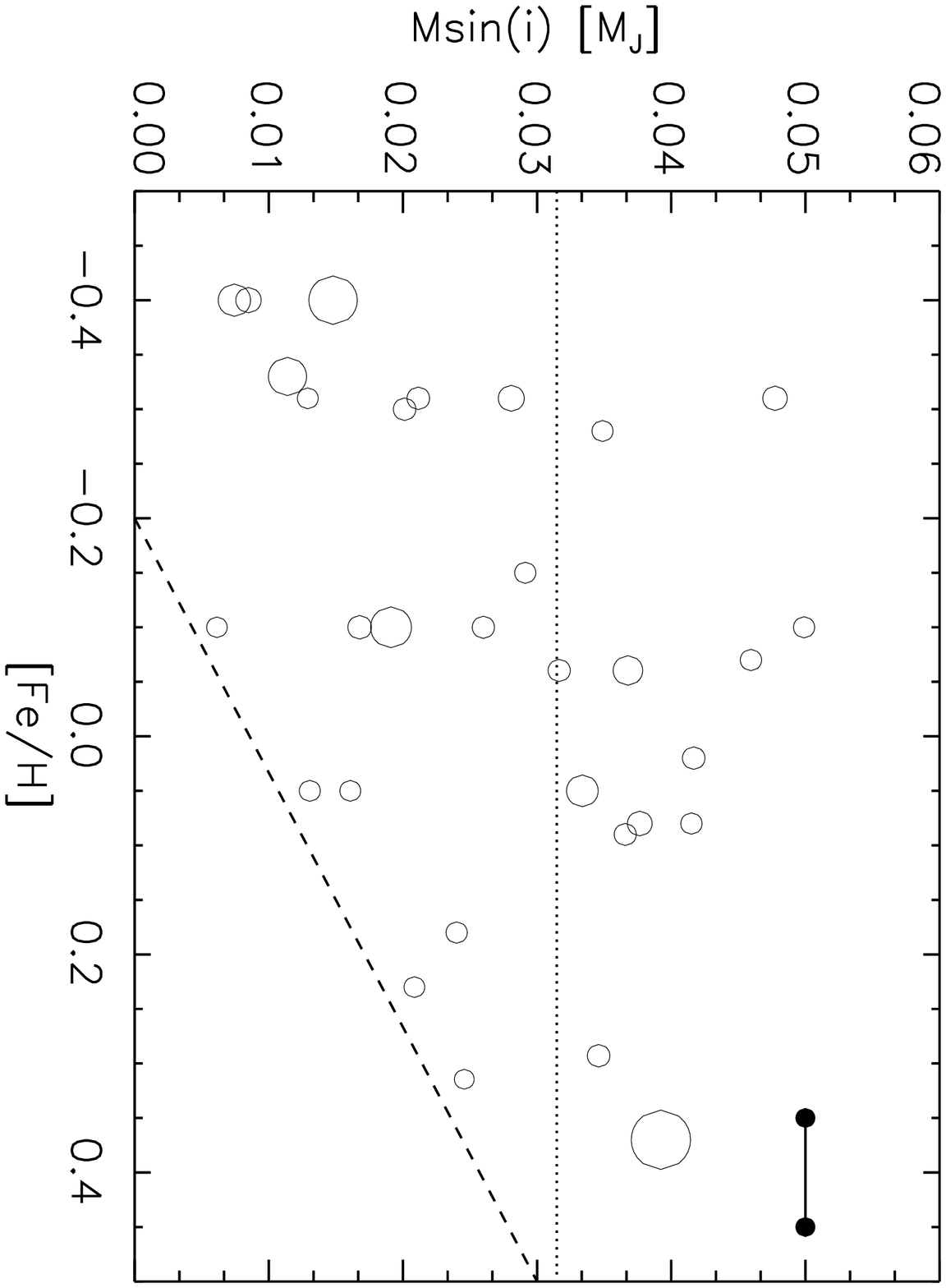}
\vspace{0cm}
\caption{Metallicity against minimum mass for all radial velocity detected planets with a minimum mass less than that of Neptune 
(\msini~$\le$~0.05~M$_{\rm{J}}$).  The horizontal dotted line highlights the canonical boundary where runaway gas accretion onto the growing core is expected to occur.  
The dashed line marks the lower boundary in mass and the data points have been scaled in size by their orbital period.  The filled circles connected by a straight line show the position 
of HD77338$b$ using our metallicity and the metallicity of \citet{trevisan11}, respectively.}
\label{rocky_met_mass}
\end{figure}

\subsection{Low-mass Systems}

Lower mass systems may show a somewhat different metallicity distribution in comparison to gas giants (\citealp{udry06}; \citealp{neves09}), with no clear metallicity bias.  
Models indicate that gas giant planets form 
outside the ice line boundary, taken to be $\sim$5~AU, in metal-poor disks (\citealp{mordasini12}) meaning smaller rocky/icy planets can form interior to this boundary, 
giving rise to the observed population of short period rocky planets around stars with a lower metal content.

Fig.~\ref{rocky_met_mass} shows the distribution of sub-Neptune mass planets (\msini$\le$0.05M$_{\rm{J}}$) as a function of metallicity using data taken from the Exoplanet 
Database\footnote{http://www.exoplanets.org} as of February 2012.  All values we have used are listed in Table~\ref{tab:planets}.  One of the most striking features we see is the 
lack of planets with high metallicities and very low masses, in the lower right corner of the plot.  Even though the numbers are relatively small at present, there 
appears a lower boundary that increases with metallicity, which would be contrary to the presumed biases in current radial velocity surveys given the number 
of metal-rich programs currently running.  HD77338 also appears to follow this trend given that it has a very high metallicity and a minimum mass commensurate with that 
of Uranus in our solar system.

A strong bias that could be manifest here is from the distribution of the orbital periods of the sample.  Given the high fraction of short period gas giants in metal-rich systems, 
it is possible that lower mass planets 
reside at longer orbital periods, meaning they will be biased against due to the well known radial velocity bias against longer period and low amplitude objects 
(see \citealp{cumming04}).  To test this we scale the data points in Fig.~\ref{rocky_met_mass} by the ratio between each planet's period and that of the longest period 
planet in the set, GJ876$e$.  This shows us that this is not the case and that as expected, the planets are mostly short period planets at the perimeter of this boundary.  
It is also noticeable that beyond the cluster of planets near the perimeter of the boundary we mark by the dashed line, there is a possible valley in the metallicity plane, then the 
rest of the sample is found.  A metallicity-mass correlation 
like this, at low-masses, would fit well with the core accretion model for planet formation, where the low-mass planets around super metal-rich stars attain a higher mass 
due to the abundance of planet forming material in the disk (\citealp{ida04b}; \citealp{mordasini12}).  What is not obvious is why there would be a lower boundary.  
Possibly the planetesimals quickly accrete material and grow much faster than at lower metallicities, meaning the boundary traces the planet formation timescales for a 
given disk metallicity.  In any case, if there is a lower boundary that holds to long period orbits for super metal-rich stars, this has consequences for the fraction of 
habitable Earth-mass planets in the galaxy, as super metal-rich stars will not be abundant in such planets.

\subsubsection{Monte Carlo Test}

To test if the lack of planets in the lower right corner of the metallicity-mass plane is statistically significant, and hence a metallicity-mass correlation for low-mass 
rocky planets is statistically significant, we perform a Monte Carlo analysis.  We setup the model in two parts where we first generate a random realisation of the sample 
of 36 low-mass planets we have tested in Fig.~\ref{rocky_met_mass}.  To generate the masses in a robust fashion we take the observationally driven mass distribution 
for lower-mass planets (\citealp{butler06}; \citealp{cumming08}).  Lopez \& Jenkins (2012, in preparation) show that a power law model such as this explains the 
observed turnover of the mass distribution at the lowest masses and hence the model is a robust representation of the current distribution of exoplanets.  
Eq$^{\rm{n}}$~\ref{eq:massfn} shows the form of the mass function we consider.

\begin{equation}
\label{eq:massfn}
dN/dM = k M^{-1.2}
\end{equation}

Here dN/dM represents the frequency of planets as a function of mass, M is the mass bin considered, and $k$ is a scaling constant to fit the observed population.  We can then 
integrate this mass function equation (Eq$^{\rm{n}}$~\ref{eq:massfnint1}) to get the result shown in Eq$^{\rm{n}}$~\ref{eq:massfnint}, and by taking the inverse we can generate 
the cumulative distribution function (CDF; Eq$^{\rm{n}}$~\ref{eq:massfncdf}) that can be used to draw masses randomly from the observed population of exoplanets.

\begin{equation}
\label{eq:massfnint1}
dN = k \int_{-\infty}^{M_{o}} M^{-1.2} dM
\end{equation}

\begin{equation}
\label{eq:massfnint}
N = \frac{-5k}{M^{0.2}}
\end{equation}

Now if we set the constant $k$ equal to -0.014 this will normalize the function such that we have a probability density function and the corresponding CDF is given by:

\begin{equation}
\label{eq:massfncdf}
CDF(M) = \left(\frac{0.069}{N} \right)^5
\end{equation}

The CDF allows us to draw random masses for planets and restrict the values to be within the range between 0.006 to 0.050~M$_{\rm{J}}$ to match our region of interest.  We can then 
randomly generate metallicities, and in this test we simply draw from a uniform distribution of [Fe/H] values in the range from -0.5 to +0.5~dex, which covers all the values 
of metallicities in our test.

Once we have built a random sample of planets we can then add the boundary and test how many times there are planets that reside under the boundary region we identify 
in Fig.~\ref{rocky_met_mass}.  We run the code 1'000'000 times to ensure a high level of robustness in the final probability measurement.  Our test reveals that for a sample 
of 36 stars, assuming a uniform metallicity distribution, 99.9993\% of the time there are planets to be found below the boundary region.  Therefore, this \emph{low-mass 
planet desert} is statistically significant at almost the $\sim$4.5$\sigma$ level, under these test conditions.  For a sample of only 
10 planets we still find that 96\% of the time there are planets below the boundary region, and for a sample of 5 planets we find a percentage of 81\%.  These tests indicate 
that there is some significant correlation between metallicity and mass for low-mass planets, at least in the metal-rich regime, that would be important to quantify in the future. 

We do note that our results can vary due to the assumed metallicity model that we use to draw the sample of random metallicities from.  For instance, the result will become more 
significant for a distribution that follows the current observed distribution for more massive exoplanets (see \citealp{sousa11}), and vice-versa for a distribution shaped 
in the opposite fashion.  

As for the possible valley, the sample is as yet too small to draw statistically significant conclusions and therefore difficult to test without detailed modeling 
of the core accretion method of forming planets, therefore we are only pointing out the possibility that there are two 
classes of low-mass planets that have a metallicity dependency.  In particular, magnitude or distance limited samples will be biased towards sub-solar metallicities 
as they will be governed by the metallicity distribution of the local galactic neighbourhood (\citealp{holmberg09}).  However, to fully probe these samples it is 
necessary to simulate the radial velocity data for each star individually (\citealp{otoole09}) to better understand the completeness of the bins we have discussed.

\subsubsection{What does Kepler Say?}

\begin{figure}
\vspace{7.0cm}
\hspace{-4.0cm}
\includegraphics{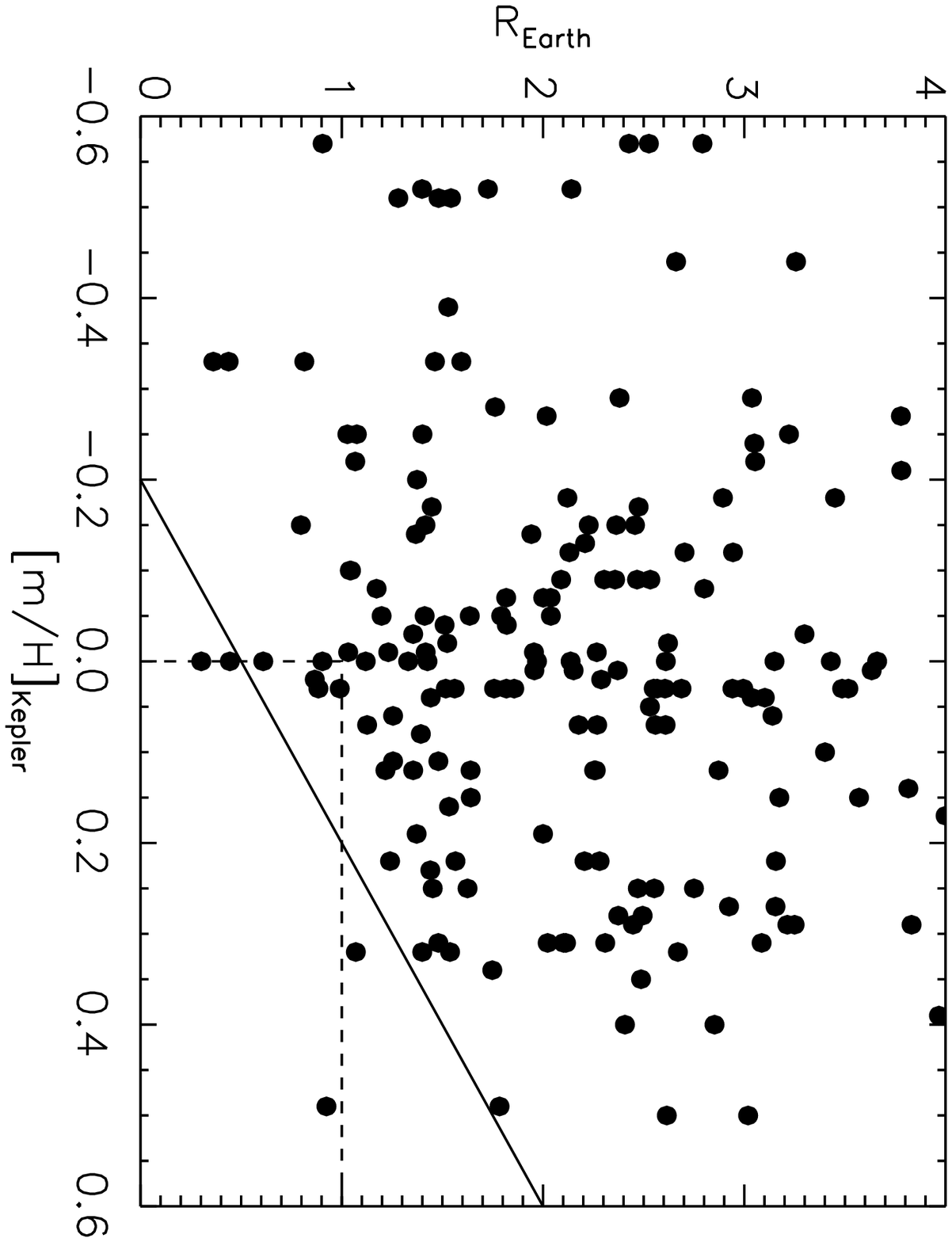}
\vspace{0cm}
\caption{Kepler results for 156 planet-hosting stars detected by Kepler in the metallicity - radius plane.  The dashed box region in the 
lower right corner highlights the lack of low-mass Kepler transits in comparison to the more metal-poor region.}
\label{kepler_met_mass}
\end{figure}

The Kepler Space Telescope has recently released a plethora of transiting planet candidates, that range in radius from below Earth radii up into the 
gas giant planet arena (\citealp{borucki11}).  Follow-up of Kepler targets are difficult due to the relative faintness of the candidate stars, however, 
\citet{buchhave12} have published the first large set of spectroscopic metallicity for 156 Kepler planet-host stars that can be used to 
draw statistically significant conclusions on Kepler host star abundance patterns.

In Fig.~\ref{kepler_met_mass} we show the Kepler results published in Buchhave et al. in the metallicity - radius plane.  We only show the stars that 
host planets with radii of 4$R_{\oplus}$ or less, and we do not distinguish between multiplanet systems and single planet host stars.  The 4$R_{\oplus}$ 
limit was chosen simply because typical mass-to-radius relationships, e.g. $M_{\rm{p}}$~=~$R_{\rm{p}}^{2.06}$ (\citealp{lissauer11}), would give rise to 
around the Neptune-mass limit highlighted in Fig.~\ref{rocky_met_mass}.  The paucity of planets we find in the metallicity - mass plane from the 
radial velocity data could therefore be manifest in the Kepler data too.  However, we note that a single mass-to-radius relationship is surely unrealistic 
due to the possible diversity of rocky planets (see \citealp{seager07}).

Studying the lower right corner of Fig.~\ref{kepler_met_mass}, the high metallicity and low mass regime, we do find a paucity of planets when we compared 
to the lower left corner, or the low metallicity and low mass region.  To better highlight this relative $planet desert$ we bound the region by a dashed box and also a 
similar rising boundary line (solid curve) as we show in the metallicity - mass plane.  The Kepler data is in good agreement with the radial velocity sample 
in respect to this paucity of planets, although not likely in absolute value, with only two planets significantly below the boundary region we highlight.  
In fact, within the boxed region, parameterised 
by the host star having a metallicity of solar or above and the transiting planet having an Earth radius or lower, there are only 8 planets, and only one of these 
planets is currently in a single system.  Therefore, it could be that really low-mass planets can exist in the super metal-rich regime if they are part of 
multi-planet systems, since under core accretion the planetesimals will compete for material to form and this will mean less material for each of the planetesimals 
to reach higher core masses.

If these features however are real, then for a given stellar/disk metallicity it appears 
that planetesimals quickly grow to their boundary region and stop there, or they then quickly make the transition to around 2 to 3 times their boundary mass and either 
continue growing to higher mass planets before the disk dissipates, or stay where they are due to depletion of the disk material.

\acknowledgments

We thank the anonymous referee for their useful suggestions and prompt response.   
We thank Rene Mendez for useful discussions on the kinematic orbits.  We also thank Claudio Melo for help in re-reducing the old HARPS data with the 
new DRS.  JSJ also acknowledges funding by Fondecyt through grant 3110004 and partial support from Centro de Astrof\'\i sica FONDAP 15010003, the  GEMINI-CONICYT FUND 
and from the Comit\'e Mixto ESO-GOBIERNO DE CHILE.  HRAJ, MT, FM, YVP, OI, and DJP are supported by RoPACS, a Marie Curie Initial Training Network 
funded by the European Commission's Seventh Framework Programme.  PR acknowledges support through Fondecyt grant 1120299.  AJ acknowledges support from 
Fondecyt project 1095213 and from the Millennium Science initiative, Chilean Ministry of Economy (Nuclei 
P10-022-F and P07-021-F).  ADJ is funded by a Fondecyt postdoctorado, under project number 3100098.  MTR received support from CATA (PB06 Conicyt).  
Based on observations made with the European Southern Observatory telescopes obtained from the ESO/ST-ECF Science Archive Facility.  We also acknowledge use of the 
Simbad and Vizier databases.

\appendix

\section{Orbital Plots for HD4308 and HD20794}

\begin{figure*}
\vspace{4.0cm}
\hspace{-4.0cm}
\includegraphics{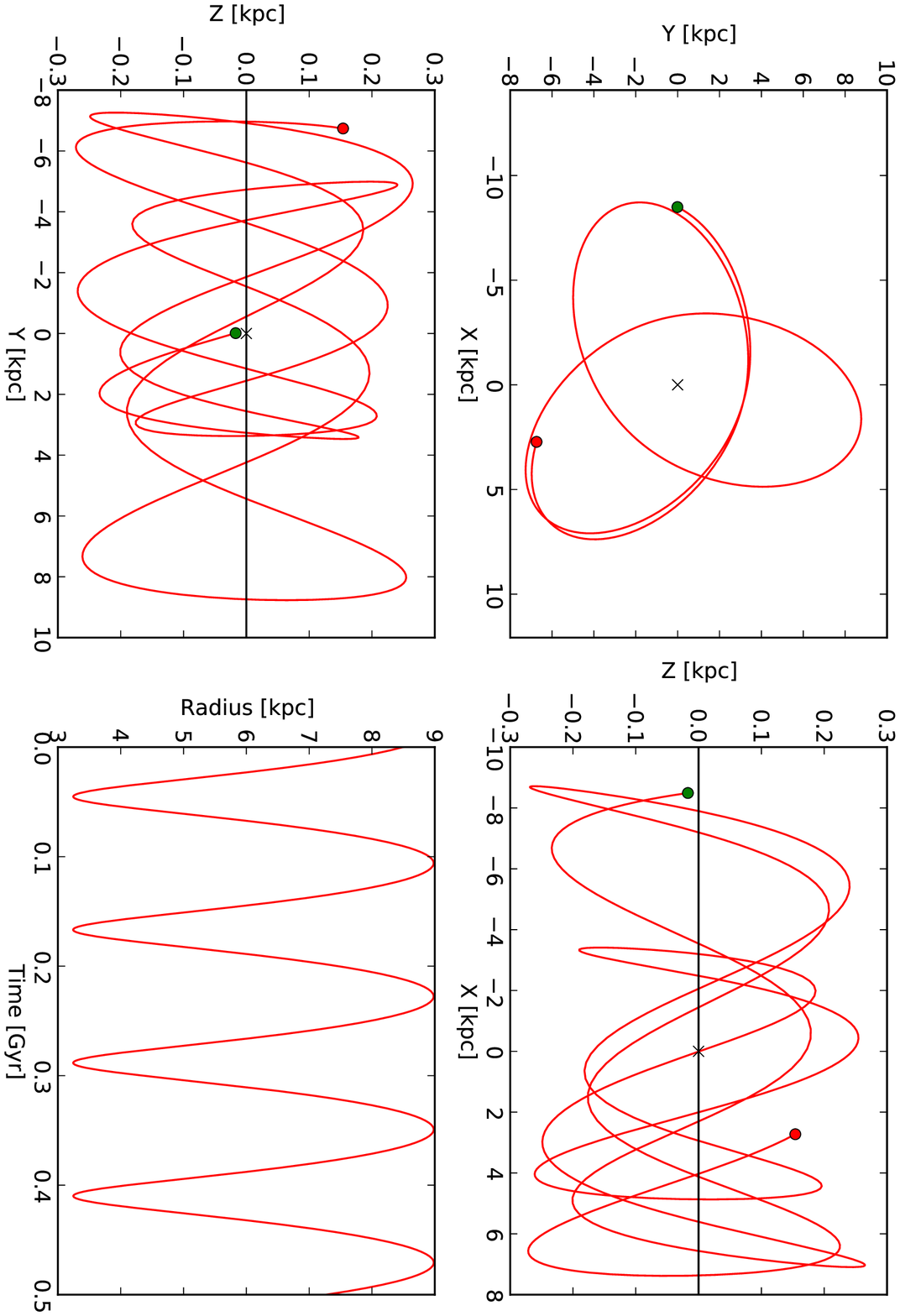}
\includegraphics{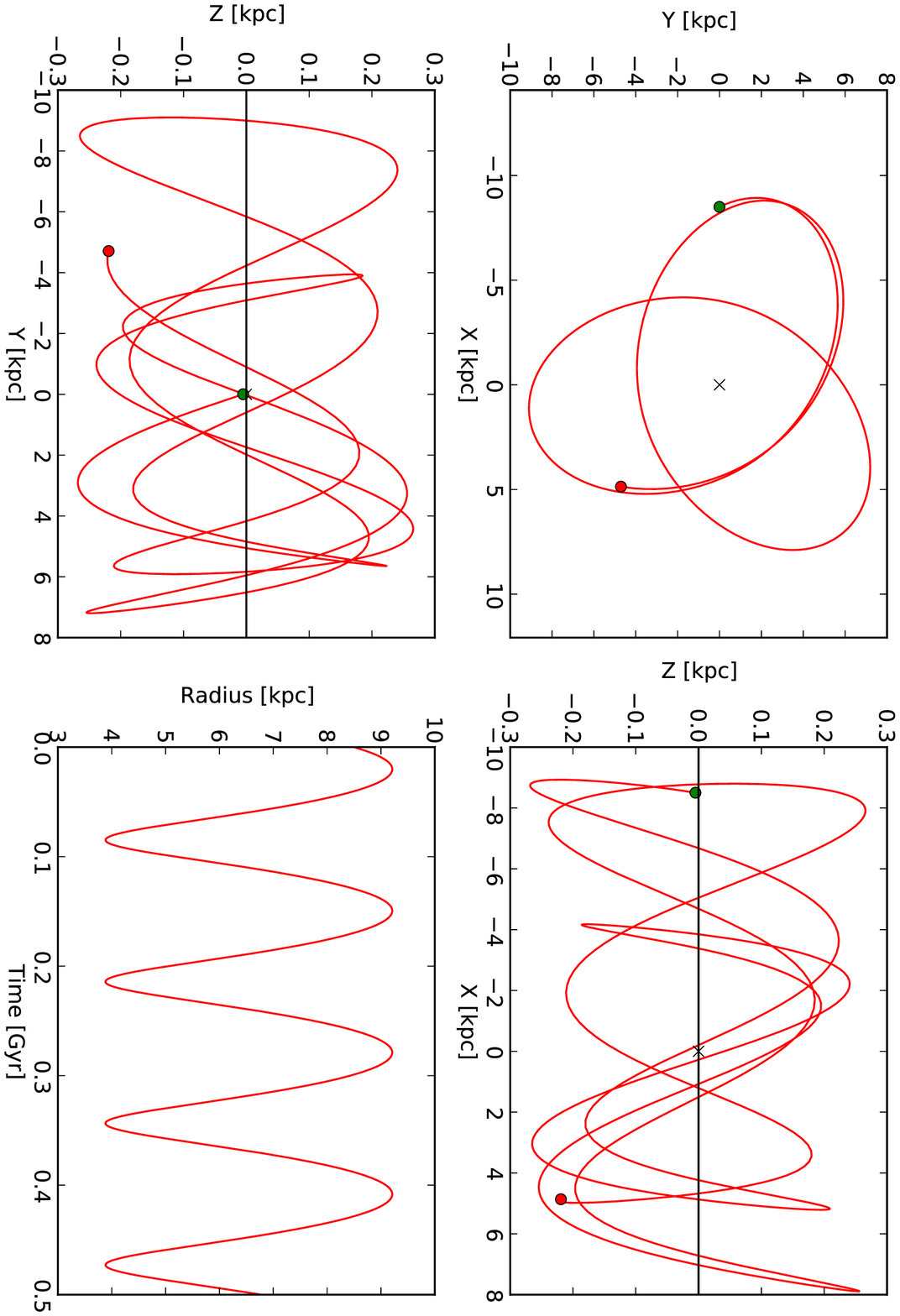}
\vspace{7.6cm}
\caption{The calculated orbital motion of HD4308 (top four panels) and HD20794 (lower four panels) through the 
galaxy.  Both show fairly eccentric orbits and are probably thick disk stars.}
\label{orbit}
\end{figure*}

\section{Orbital Plots for HD77338 and the Sun}

\begin{figure*}
\vspace{4.0cm}
\hspace{-4.0cm}
\includegraphics{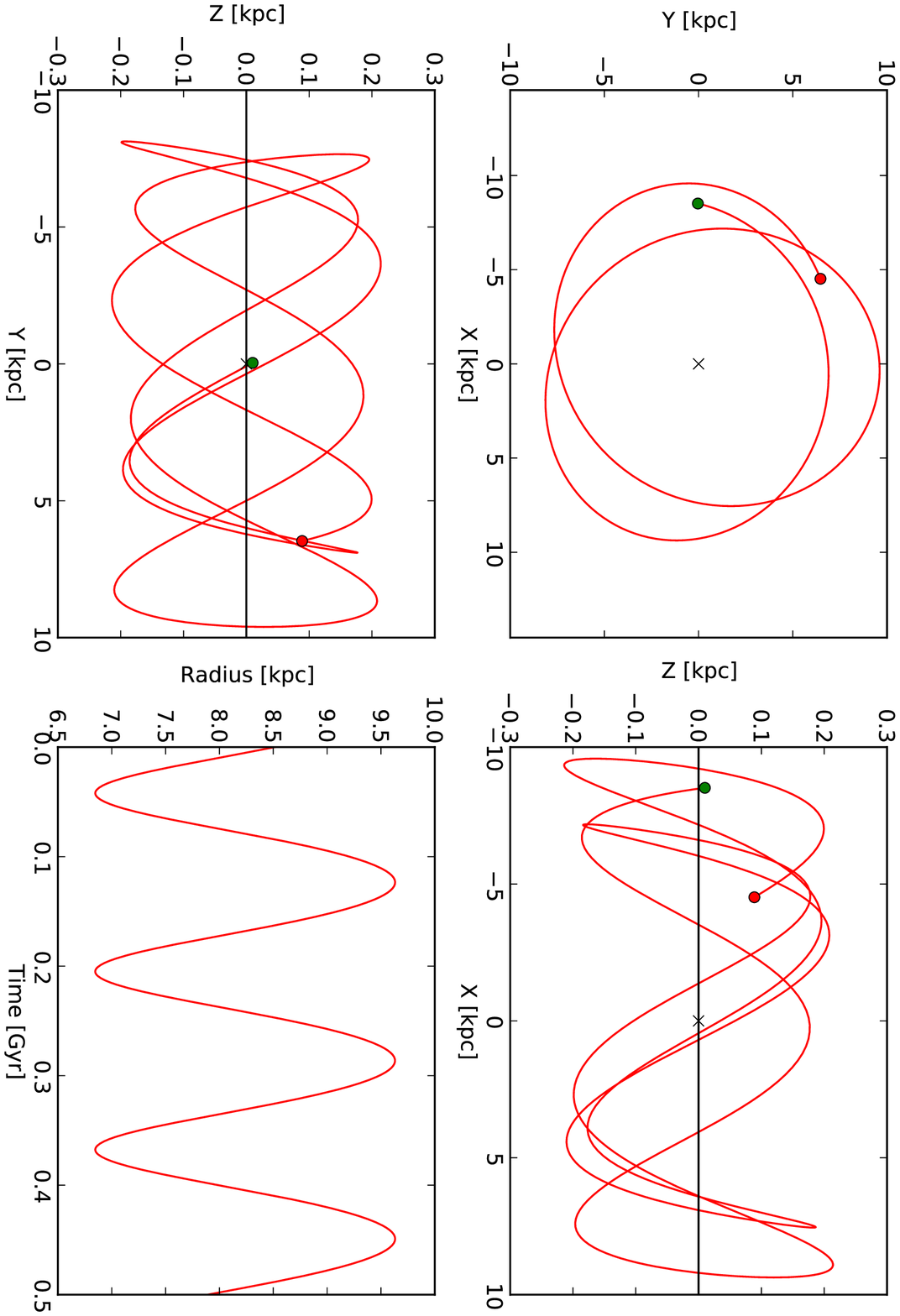}
\includegraphics{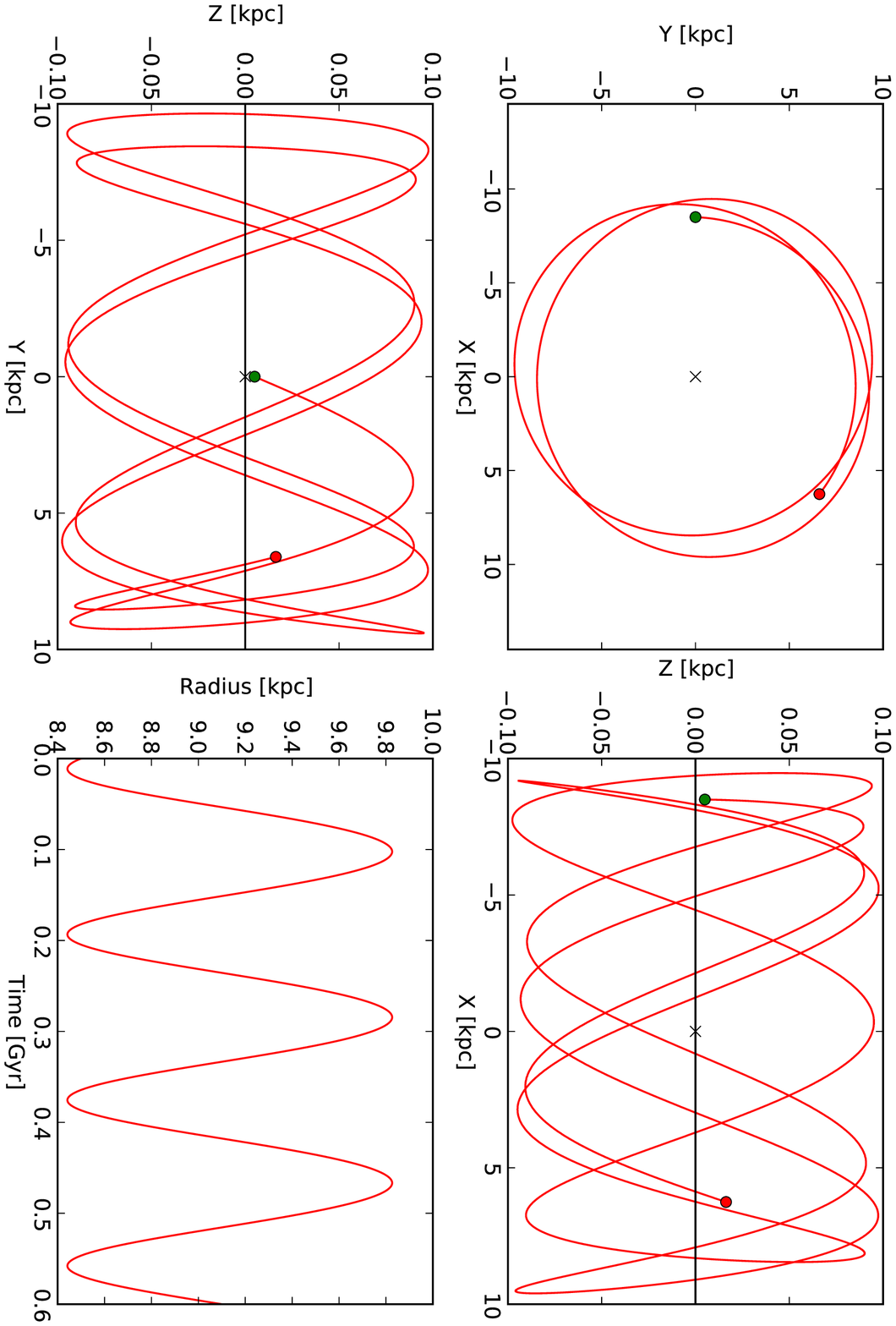}
\vspace{7.6cm}
\caption{The calculated orbital motion of HD77338 through the galaxy in all three spatial dimensions X, Y, Z, along with 
the distance from the galactic center are show as a function of time are shown in the top four panels, from top left to lower 
right, respectively.  The same orbital calculations are shown in the lower four panels for the orbit of the Sun around the 
galaxy.}
\label{orbit}
\end{figure*}

\bibliographystyle{aa}
\bibliography{Jenkins_2012b}

\end{document}